\documentclass[12pt]{article}
\usepackage[T1]{fontenc}
\usepackage[utf8]{inputenc}
\usepackage{url}
\usepackage{amsbsy,amsfonts,amsmath,amssymb,amsthm,enumerate,graphicx,rotating}
\usepackage[authoryear]{natbib}
\makeatletter
\providecommand\@bibsetup[1]{}
\makeatother
\newcommand{\affil}[1]{\\ {\normalsize #1}}
\newenvironment{keywords}{\noindent\textbf{Keywords:} }{\par}

\title{Extrema, Barrier Options, and Semi-Analytic Leverage Corrections in Stochastic-Clock Volatility Models}

\author{Tristan Guillaume$^{\ast}$$\dag$\thanks{$^\ast$Corresponding author. Email: tristan.guillaume@cyu.fr}\\
\affil{$\dag$CY Cergy Paris Universit\'e, Laboratoire Thema, 33 boulevard du port, F-95011 Cergy-Pontoise Cedex, France}}

\begin{document}
\maketitle

\begin{abstract}
Barrier derivatives depend on extrema and first-passage events and are therefore highly sensitive to volatility dynamics---especially to the instantaneous return--volatility correlation $\rho$, often called ``leverage''. This sensitivity makes accurate and fast pricing under realistic stochastic-volatility specifications difficult: two-dimensional PDE solvers are expensive inside calibration loops, while Monte Carlo methods converge slowly when barrier hits are rare and discretely monitored. In equity markets in particular, the pronounced implied-volatility skew motivates factoring in a negative return--volatility correlation.

We study a class of continuous-path stochastic-clock volatility models in which the log-price is represented as a Brownian motion run on a random increasing clock. In the baseline independent-clock case ($\rho=0$), a broad family of barrier-relevant objects---maximum distributions, survival probabilities, and killed joint laws---reduces to one-dimensional quantities determined by the Laplace transform of the terminal clock. This yields transform-only pricing formulas for single- and double-barrier contracts that are fast and numerically stable once the clock transform is available, notably for affine and quadratic clocks.

To incorporate leverage without forfeiting tractability, we develop a systematic small-$\rho$ expansion around the $\rho=0$ backbone. The expansion produces a hierarchy of forced problems whose forcing terms are semi-analytic and computable from baseline barrier objects. We provide two implementable leverage-correction routes: forced PDEs and a Duhamel-type Monte Carlo representation, and we show how Padé acceleration can extend practical accuracy to equity-like correlations. Calibration then proceeds by: (i) fitting clock parameters from vanillas using only one-dimensional transforms, (ii) precomputing the $\rho=0$ barrier backbone once, and (iii) iterating on $\rho$ (and any remaining parameters) using the fast semi-analytic corrections---optionally Padé-accelerated---inside a standard least-squares loop.
\end{abstract}

\begin{keywords}
stochastic clock; time change; barrier options; maximum distribution; killed diffusion; Laplace transform; leverage effect; implied-volatility skew; small-correlation expansion; Padé approximation; calibration
\end{keywords}


\section{Introduction}

Barrier derivatives are ubiquitous in equity and foreign exchange markets. Unlike European options, whose values depend only on the terminal distribution of the underlying, barrier contracts are sensitive to the entire trajectory: their payoffs are contingent on whether the asset price crosses specified threshold levels during the contract's life. This path dependence concentrates model risk in regions that are weakly constrained by vanilla calibration. Two stochastic volatility models that produce indistinguishable implied volatility surfaces can generate materially different barrier prices, because they imply different hitting probabilities and different joint laws of the terminal value and the running maximum. The practical consequence is well known to practitioners: barrier options are notoriously difficult to hedge and price reliably, and model misspecification can lead to substantial profit-and-loss volatility. Classical discussions of barrier payoffs, monitoring effects, and replication challenges include Reiner and Rubinstein (1991) and Broadie et al. (1997).

From a computational standpoint, the same path dependence that creates model risk also concentrates computational cost precisely where calibration requires speed. Under stochastic volatility with leverage---the empirically dominant case in equity markets, where instantaneous returns and volatility shocks are negatively correlated---barrier pricing typically demands either (i) repeated two-dimensional partial differential equation (PDE) solves with mixed-derivative structure, as required by classical leverage models such as Heston (1993), or (ii) Monte Carlo procedures whose variance deteriorates for rare barrier events and whose bias depends sensitively on monitoring frequency and bridge corrections (Broadie et al., 1997; Glasserman, 2003). Existing numerical methodologies for barriers under stochastic volatility are substantial (see, e.g., (Gobet, 2000; Jeon et al., 2017; Guterding and Boenkost, 2018)), yet for applied work there remains a persistent gap between models rich enough to reproduce leverage-induced skew and procedures robust and fast enough for calibration loops that include even a modest set of barrier quotes.

\subsection{A stochastic-clock perspective}

A large family of continuous-path stochastic volatility models can be expressed---exactly or to high accuracy---as a time-changed Brownian motion, a representation rooted in Bochner's theory of subordination (Bochner, 1949) and developed extensively in financial applications (Clark, 1973; Carr and Wu, 2004; Carr et al., 2003). In this framework, the log-price process $\left\{X_{t},t\geq 0\right\}$ under a risk-neutral measure takes the form

\begin{equation}
X_{t}=x_{0}+\beta\Gamma_{t}+B_{\Gamma_{t}}
\tag{1.1}\label{eq:1.1}
\end{equation}

where B is a standard Brownian motion and $\Gamma$ is an almost surely increasing random process---the stochastic clock---typically interpreted as cumulative variance: $\Gamma_{t}:=\int_{0}^{t} v_{s}ds$, for an instantaneous variance process ($v_{t}$). The clock governs how quickly randomness is consumed: large increments of $\Gamma$ correspond to volatile periods. This representation unifies a broad class of specifications, including integrated CIR (Heston-type) and squared Ornstein--Uhlenbeck variance processes, as well as more general affine, quadratic-Gaussian, and Wishart clock families (Cox et al., 1985; Duffie et al., 2000; Cuchiero et al., 2011).

The central observation for barrier applications is that many path functionals of X can be computed by conditioning on the terminal clock $\Gamma_{t}$. When the clock is independent of the Brownian driver ---the case $\rho = 0$ in the standard two-factor stochastic volatility parameterization---the conditional law of X given $\Gamma_{t}$ is Gaussian, and conditional extrema reduce to classical Brownian reflection identities (Borodin and Salminen, 2002). Unconditioning then expresses barrier and killed-law objects as one-dimensional transforms in the distribution of $\Gamma_{t}$ and hence through the Laplace transform

\begin{equation}
\Phi_{T}(\lambda):=\mathbb{E}\left[e^{-\lambda\Gamma_{T}}\right], \lambda \geq 0
\tag{1.2}\label{eq:1.2}
\end{equation}

This turns barrier pricing into a transform-only exercise: once $\Phi_{T}$ is available---explicitly for affine and quadratic clocks, or via low-dimensional Riccati systems---the computational complexity becomes comparable to Fourier-based pricing for vanillas (Carr and Madan, 1999; Fang and Oosterlee, 2009).

\subsection{The leverage problem and its resolution}

While the independent-clock framework is analytically powerful, it is generally insufficient to reproduce equity-like implied volatility skews, which are driven in large part by leverage---nonzero instantaneous correlation $\rho$ between the return driver and the variance/clock driver. The empirical magnitude of this correlation, typically $\rho\approx -0.7$ for equity indices (Gatheral, 2006), is far from negligible. When $\rho\neq 0$, the time-change reduction fails: the Brownian motion driving returns is no longer independent of the clock, the conditional Gaussian structure is lost, and barrier problems become genuinely two-dimensional. Numerically, this is precisely the regime where repeated mixed-derivative PDE solves---or heavy Monte Carlo---make barrier-inclusive calibration expensive and brittle.

The core methodological contribution of this paper is a leverage layer that retains the independent-clock case as a solvable anchor and incorporates correlation through a systematic expansion around $\rho = 0$. Concretely, if $u(\rho)$ denotes a barrier-relevant functional (price, survival probability, or killed-law quantity), we develop an approximation of the form

\begin{equation}
u(\rho) \approx u_{0}+\rho u_{1}+\rho^{2}u_{2}+\cdots,
\tag{1.3}\label{eq:1.3}
\end{equation}

where $u_{0}$ corresponds to the independent-clock case and the correction coefficients $(u_{k})_{k\geq 1}$ are obtained from a forced hierarchy whose forcing terms are semi-analytic and expressed through baseline ($\rho = 0$) barrier objects. The key structural insight is that correlation enters the two-factor generator linearly through a single mixed-derivative operator, so that all correction coefficients solve killed boundary problems under the same decoupled generator $\mathcal{L}_{0}$ that governs the baseline---only the forcing term changes. This avoids the expensive mixed-derivative structure of the full $\rho\neq 0$ problem at each order of the expansion.

\subsection{Relation to prior literature}

The barrier-option literature under stochastic volatility is extensive, and we position our work relative to three strands.

(a) Independent time-change barrier theory.

The reduction of barrier quantities to Laplace transforms of the clock under independence is well established. Hieber and Scherer (2012) develop first-passage and double-barrier exit quantities for continuously time-changed Brownian motion, and Escobar et al. (2014) provide efficient series pricing for double-barrier derivatives. These works remain the closest structural anchors for our baseline ($\rho = 0$) layer. The present paper does not claim novelty for the existence of spectral series or integral representations under independent time changes; rather, we provide a unified, implementation-ready interface that integrates these representations with a broad menu of tractable clocks and prepares them as the zeroth-order building block for leverage corrections.

(b) Robust and semi-robust barrier replication.

The put--call symmetry program of Carr and Lee (2009) and subsequent robust replication work address barrier-like claims through model-free or semi-model-free identities, typically requiring access to a continuum of vanilla prices and imposing symmetry or structural assumptions. Our approach is complementary: it is explicitly model-based, but in exchange produces direct formulas for distributions of extrema and barrier prices from $\Phi_{T}$, and extends naturally to clock families beyond the classic affine set.

(c) Correlated stochastic volatility barriers.

Most post-2014 work on barriers under correlated stochastic volatility relies on heavier machinery: full two-dimensional PDEs with mixed derivatives (Guterding and Boenkost, 2018; Li and Zhang, 2016), integral equation methods (Jeon et al., 2017), broad stochastic diffusive volatility combined with jumps (Guardasoni and Sanfelici, 2016; Kirkby et al., 2017). These are important contributions, but they do not preserve the plug-in-$\Phi_{T}$-and-compute structure that makes the independent-clock case powerful. Our $\rho$-expansion is designed precisely to fill this gap: it keeps the exact one-dimensional clock formulas as the leading term and adds leverage through forced equations without mixed derivatives, yielding a semi-analytic correction that is computationally light enough to embed in calibration.

\subsection{Contributions and novelty}

The paper develops a coherent methodology that makes a barrier-capable stochastic-clock framework both calibratable and implementable. The principal contributions are:

(a) Transform-only framework for extrema and killed laws under an independent continuous clock, and double-barrier pricing via a rapidly convergent series with explicit coefficients

By conditioning on $\Gamma_{t}$ and averaging using $\Phi_{T}$, we obtain explicit representations for maximum distributions, survival probabilities, and killed joint laws, yielding fast building blocks for single-barrier claims expressed as single real integrals involving only elementary functions and the clock Laplace transform. We also derive a real sine series (Dirichlet spectral expansion) for corridor survival and double knock-out payoffs, with coefficients written in closed form and model dependence entering solely through $\Phi_{T}$ evaluated on a discrete positive grid. This enables efficient caching and reuse across strikes, payoffs, and calibration iterations.

(b) A calibration-feasible leverage layer: the $\rho$-expansion with semi-analytic forcing

The forced hierarchy avoids repeated full mixed-derivative PDE solves and admits two practical implementation routes: a forced-PDE route and a Duhamel-type Monte Carlo route, both operating under the decoupled $\rho = 0$ dynamics. The critical enabler is that the forcing terms at each order are themselves semi-analytic, constructed by differentiating the baseline barrier integrals with respect to the clock state---a differentiate-the-transform mechanism that sidesteps noisy finite differencing near barriers.

(c) Padé acceleration with stability diagnostics

Raw Taylor truncation of the $\rho$-expansion is accurate only for small $|\rho|$. To extend the method to equity-like correlations ($|\rho|\approx 0.5$--$0.9$), we develop Padé resummation with explicit pole tests. Numerical experiments show that the combination of fifth-order expansion and [2/2] or [3/2] Padé approximants achieves errors below 0.3\% even at $|\rho|=0.9$.

(d) Calibration workflow aligned with practitioner constraints

We propose a structured workflow: (i) fit the clock parameters to vanillas using fast transform pricing; (ii) incorporate a small barrier set to pin down path-dependent behaviour; (iii) calibrate leverage via cached expansion coefficients, scanning over $\rho$ at negligible marginal cost---together with explicit error diagnostics and optional Brownian-bridge corrections for discrete monitoring (Broadie et al., 1997).

\subsection{Organization of the paper}

The remainder of the paper is organized as follows.

Section 2 develops the transform-only valuation framework for single-barrier options under an independent stochastic clock. We derive the killed terminal density (Proposition 2.1), reduce the joint survival CDF to a single real integral (Proposition 2.2), and obtain explicit clock-modular formulas for the up-and-out put (Theorem 2.1) and the down-and-out call (Theorem 2.2).

Section 3 extends the framework to double knock-out options, deriving a rapidly convergent real sine series whose coefficients depend on the clock only through Laplace transform evaluations on a discrete positive grid (Theorem 3.1).

Section 4 catalogs solvable stochastic clocks---integrated CIR, squared Ornstein--Uhlenbeck, multi-factor quadratic-Gaussian, Markov-switching, and affine jump-variance specifications---providing explicit or low-dimensional Riccati-based expressions for the Laplace transform interface required by Sections 2--3.

Section 5 introduces the leverage layer. We decompose the correlated two-factor generator as $\mathcal{L}_{\rho}=\mathcal{L}_{0}+\rho\mathcal{L}_{1}$, derive the $\rho$-expansion and its forced hierarchy (Proposition 5.1), and discuss semi-analytic differentiation of the baseline barrier integrals for constructing forcing terms.

Section 6 presents numerical experiments. We compare results using the independent-clock formulas against Monte Carlo for CIR and squared OU clocks under two volatility regimes and two maturities, then systematically evaluate the $\rho$-expansion and Padé acceleration for correlations ranging from -0.9 to +0.9.

Section 7 describes the calibration workflow, including initialization from variance swaps and ATM volatilities, joint vanilla-plus-barrier fitting, leverage calibration via cached expansion coefficients, and practitioner-oriented diagnostics for detecting extrapolation beyond the expansion's reliable regime.

\section{Single-barrier valuation under an independent stochastic clock: exact one-dimensional real-integral formulas}

This section derives clock-modular valuation formulas for two canonical continuously monitored single-barrier contracts: the up-and-out put (UOP) and the down-and-out call (DOC). Under an independent stochastic clock, pricing reduces to real-axis integral representations involving only elementary functions and the real-axis Laplace transform of the terminal clock.

\subsection{Pricing model and standing assumptions}

Let $(S_{t})_{0\leq t\leq T}$ be the spot price of a traded underlying with constant, deterministic short rate $r$ and dividend yield $q$. Under a pricing measure $Q$ associated with the money-market numeraire,

\begin{equation}
\frac{dS_{t}}{S_{t}}=(r-q)dt+\sqrt{v_{t}}\,dW_{t}^{Q}, \qquad 0\leq t\leq T
\tag{2.1}\label{eq:2.1}
\end{equation}

Fix maturity $T$ and introduce the $T$-forward (carry-adjusted) price

\begin{equation}
F_{t}:=S_{t}e^{(r-q)(T-t)}
\tag{2.2}\label{eq:2.2}
\end{equation}

so that $F_{T}=S_{T}$. The time-0 value of any $F_{T}$-measurable payoff $\Pi$ may be written under the $T$-forward measure $\mathbb{Q}^{T}$ as

\begin{equation}
V_{0}=P(0,T)\,E^{\mathbb{Q}^{T}}[\Pi], \qquad P(0,T)=e^{-rT}
\tag{2.3}\label{eq:2.3}
\end{equation}

Assume that under $\mathbb{Q}^{T}$,

\begin{equation}
\frac{dF_{t}}{F_{t}}=\sqrt{v_{t}}dW_{t}
\tag{2.4}\label{eq:2.4}
\end{equation}

where $W$ is a $\mathbb{Q}^{T}$-Brownian motion and $v_{t}\geq 0$ is an adapted activity rate. Define the clock

\begin{equation}
\Gamma_{t}:=\int_{0}^{t} v_{s}ds
\tag{2.5}\label{eq:2.5}
\end{equation}

and set

\begin{equation}
X_{t}:=\log F_{t}, \qquad x_{0}:=\log F_{0}
\tag{2.6}\label{eq:2.6}
\end{equation}

Then

\begin{equation}
dX_{t}=-\frac{1}{2}v_{t}dt+\sqrt{v_{t}}dW_{t}
\tag{2.7}\label{eq:2.7}
\end{equation}

\noindent\textbf{Assumption 2.1 (independent continuous clock).} The process $v$ (equivalently $\Gamma$) is independent of $W$, and $\Gamma$ is almost surely continuous increasing with $\Gamma_{T}<\infty$.

Under Assumption 2.1, the martingale term in \eqref{eq:2.7} has quadratic variation $\Gamma_{t}$. By Dambis--Dubins--Schwarz, there exists a Brownian motion $B$, independent of $\Gamma$, such that

\begin{equation}
\int_{0}^{t} \sqrt{v_{s}}dW_{s}=B_{\Gamma_{t}}
\tag{2.8}\label{eq:2.8}
\end{equation}

Consequently,

\begin{equation}
X_{t}=x_{0}+\beta\Gamma_{t}+B_{\Gamma_{t}}, \beta:=-\frac{1}{2}
\tag{2.9}\label{eq:2.9}
\end{equation}

The model dependence of the independent-clock regime enters through the Laplace transform of $\Gamma_{T}$

\begin{equation}
\Phi_{T}(\lambda):=E^{\mathbb{Q}^{T}}\left[e^{-\lambda\Gamma_{T}}\right], \lambda\geq 0
\tag{2.10}\label{eq:2.10}
\end{equation}

Define $\tau_{A}$ as the exit time of set $A$ by the process $X$:

\begin{equation}
\tau_{A}:=\inf\{t\geq 0: X_{t}\notin A\}
\tag{2.11}\label{eq:2.11}
\end{equation}

Condition on $\Gamma_{T}=g$. Let $Y$ be the operational-time drifted Brownian motion:

\begin{equation}
Y_{s}:=x_{0}+\beta s+B_{s}, \qquad 0\leq s\leq g
\tag{2.12}\label{eq:2.12}
\end{equation}

and define its exit time from set $A$ by

\begin{equation}
\sigma_{A}:=\inf\{s\geq 0: Y_{s}\notin A\}
\tag{2.13}\label{eq:2.13}
\end{equation}

On the event $\{\Gamma_{T}=g\}$ with $g>0$, the continuity and monotonicity of $\Gamma$ imply $\Gamma([0,T])=[0,g]$, and since $X_{t}=Y_{\Gamma_{t}}$ we obtain the equality

\begin{equation}
\{\tau_{A}>T\}=\{\sigma_{A}>g\}, \qquad X_{T}=Y_{g}
\tag{2.14}\label{eq:2.14}
\end{equation}

\subsection{Killed density for the upper barrier survival}

Fix an upper log-barrier $h\in \mathbb{R}$ with $x_{0}<h$, and define the running maximum and the survival event

\begin{equation}
M_{T}:=sup_{0\leq t\leq T}X_{t}, {\tau_{h}>T}={M_{T}<h}
\tag{2.15}\label{eq:2.15}
\end{equation}

Condition on $\Gamma_{T}=g$. By \eqref{eq:2.14}, $(X_{T},M_{T})$ has the same law as $\left(Y_{g},sup_{0\leq s\leq g}Y_{s}\right)$.

Let

\begin{equation}
p_{\beta}(g;x_{0},x):=\frac{1}{\sqrt{2\pi g}}\exp\left(-\frac{(x-x_{0}-\beta g)^{2}}{2g}\right)
\tag{2.16}\label{eq:2.16}
\end{equation}

be the transition density of $Y_{g}$ started from $x_{0}$.

By the classical method of images (or, equivalently, using the principle of reflection), the killed transition density $p_{\beta}^{(h)}(g;x_{0},x)$ on $(-\infty,h)$ is given by:

\begin{equation}
p_{\beta}^{(h)}(g;x_{0},x)=p_{\beta}(g;x_{0},x)-e^{2\beta(h-x_{0})}p_{\beta}\left(g;2h-x_{0},x\right), x<h
\tag{2.17}\label{eq:2.17}
\end{equation}

Hence, for $x<h$,

\begin{equation}
\mathbb{Q}^{T}\left(X_{T}\in dx, M_{T}<h\mid\Gamma_{T}=g\right)=p_{\beta}^{(h)}(g;x_{0},x)dx
\tag{2.18}\label{eq:2.18}
\end{equation}

Unconditioning yields the killed terminal density

\begin{equation}
j_{\beta}^{(h)}(x):=\frac{d}{dx}\mathbb{Q}^{T}(X_{T}\leq x, M_{T}<h)=E^{\mathbb{Q}^{T}}\left[p_{\beta}^{(h)}(\Gamma_{T};x_{0},x)\right], x<h
\tag{2.19}\label{eq:2.19}
\end{equation}

A key advantage of the independent-clock setting is that $f_{\beta}^{(h)}$ can be expressed using only $\Phi_{\Gamma}$ at real arguments.

\noindent\textbf{Proposition 2.1 (killed density).} Under Assumption 2.1 and a mild integrability condition (e.g. $E^{\mathbb{Q}^{T}}[\Gamma_{T}^{-1/2}]<\infty$), for every $x<h$,

\begin{equation}
j_{\beta}^{(h)}(x)=\frac{2}{\pi}e^{\beta(x-x_{0})}\int_{0}^{\infty} \sin\left(u(h-x_{0})\right)\sin\left(u(h-x)\right)\Phi_{\Gamma}\left(\frac{u^{2}+\beta^{2}}{2}\right)du
\tag{2.20}\label{eq:2.20}
\end{equation}

\begin{proof}

First consider $\beta=0$. The method of images gives the driftless killed density on $(-\infty,h)$,

\begin{equation}
p_{0}^{(h)}(g;x_{0},x)=p_{0}(g;x_{0},x)-p_{0}(g;2h-x_{0},x), x<h.
\tag{2.21}\label{eq:2.21}
\end{equation}

Applying the Fourier--cosine representation of the Gaussian kernel to \eqref{eq:2.21} yields

\begin{equation}
p_{0}^{(h)}(g;x_{0},x)=\frac{1}{\pi}\int_{0}^{\infty} e^{-(u^{2}/2)g}\left(\cos(u(x-x_{0}))-\cos(u(x-(2h-x_{0})))\right)du
\tag{2.22}\label{eq:2.22}
\end{equation}

\begin{equation}
=\frac{2}{\pi}\int_{0}^{\infty} e^{-(u^{2}/2)g}\sin(u(h-x_{0}))\sin(u(h-x))du, x<h
\tag{2.23}\label{eq:2.23}
\end{equation}

Now take general $\beta\in \mathbb{R}$. The killed density $p_{\beta}^{(h)}(\cdot;x_{0},\cdot)$, viewed as a function of $(g,x)\in(0,\infty)\times(-\infty,h)$, satisfies the forward Kolmogorov equation:

\begin{equation}
\partial_{g}p_{\beta}^{\left(h\right)}\left(g;x_{0},x\right)=-\beta\partial_{x}p_{\beta}^{\left(h\right)}\left(g;x_{0},x\right)+\frac{1}{2}\partial_{xx}p_{\beta}^{\left(h\right)}\left(g;x_{0},x\right)
\tag{2.24}\label{eq:2.24}
\end{equation}

with Dirichlet boundary: $p_{\beta}^{\left(h\right)}\left(g;x_{0},h\right)=0$ and initial condition $p_{\beta}^{(h)}(0;x_{0},\cdot)=\delta_{x_{0}}$ on $(-\infty,h)$ in the sense of distributions. It is easy to verify that

\begin{equation}
p_{\beta}^{(h)}(g;x_{0},x)=\exp\left(\beta(x-x_{0})-\frac{1}{2}\beta^{2}g\right)p_{0}^{(h)}(g;x_{0},x), x<h
\tag{2.25}\label{eq:2.25}
\end{equation}

Combining \eqref{eq:2.25} with \eqref{eq:2.23} yields

\begin{equation}
p_{\beta}^{(h)}(g;x_{0},x)=\frac{2}{\pi}e^{\beta(x-x_{0})}\int_{0}^{\infty} e^{-\frac{1}{2}(u^{2}+\beta^{2})g}\sin(u(h-x_{0}))\sin(u(h-x))du
\tag{2.26}\label{eq:2.26}
\end{equation}

It only remains to insert \eqref{eq:2.26} into \eqref{eq:2.19} and justify exchanging expectation and the $u$-integral:

\begin{equation}
\int_{0}^{\infty} e^{-\frac{1}{2}(u^{2}+\beta^{2})g}\sin(u(h-x_{0}))\sin(u(h-x))du\leq\int_{0}^{\infty} e^{-\frac{1}{2}(u^{2}+\beta^{2})g}du
\tag{2.27}\label{eq:2.27}
\end{equation}
\[\int_{0}^{\infty} e^{-\frac{1}{2}(u^{2}+\beta^{2})g}du\leq\int_{0}^{\infty} e^{-(u^{2}/2)g}du=\sqrt{\frac{\pi}{2g}}\]

so the $u$-integral (as a function of $g$) is dominated by a constant multiple of $g^{-1/2}$. Under $E^{\mathbb{Q}^{T}}[\Gamma_{T}^{-1/2}]<\infty$, Fubini's theorem applies and thus yields

\begin{equation}
j_{\beta}^{(h)}(x)=\frac{2}{\pi}e^{\beta(x-x_{0})}\int_{0}^{\infty} \sin(u(h-x_{0}))\sin(u(h-x))E^{\mathbb{Q}^{T}}\left[e^{-\frac{1}{2}(u^{2}+\beta^{2})\Gamma_{T}}\right]du
\tag{2.28}\label{eq:2.28}
\end{equation}

which, using \eqref{eq:2.10}, is exactly \eqref{eq:2.20}.

\end{proof}

\subsection{The joint survival CDF as a single real integral}

For $x<h$, define

\begin{equation}
J(x;h):=\mathbb{Q}^{T}\left(X_{T}\leq x, M_{T}<h\right)
\tag{2.29}\label{eq:2.29}
\end{equation}

\noindent\textbf{Proposition 2.2 (single real integral for the joint CDF).} Under Assumption 2.1, for every $x<h$,

\begin{equation}
J_{\beta}(x;h)=\frac{2}{\pi}e^{\beta(x-x_{0})}\int_{0}^{\infty} \sin\left(u(h-x_{0})\right)\frac{u\cos\left(u(h-x)\right)-\beta\sin\left(u(h-x)\right)}{u^{2}+\beta^{2}}\Phi_{T}\left(\frac{u^{2}+\beta^{2}}{2}\right)du
\tag{2.30}\label{eq:2.30}
\end{equation}

\begin{proof}

By definition of $J_{\beta}(x;h)$ and $j_{\beta}^{(h)}(x)$,

\begin{equation}
J_{\beta}(x;h)=\int_{-\infty}^{x} j_{\beta}^{(h)}(y)dy
\tag{2.31}\label{eq:2.31}
\end{equation}

Insert \eqref{eq:2.20} in \eqref{eq:2.31} and interchange the order of integration under the same domination/Fubini argument as the one used in Proposition 2.1 (hence under the same mild integrability condition $E^{\mathbb{Q}^{T}}[\Gamma_{T}^{-1/2}]<\infty$). The resulting inner integral is elementary; substitution yields \eqref{eq:2.30}.

\end{proof}

\subsection{Up-and-out put: single real integral valuation formula}

Let the (constant) forward upper barrier be $H>0$ and strike $K>0$, with $F_{0}<H$. Set

\begin{equation}
h:=\log H, \qquad k:=\log K, \qquad x_{0}:=\log F_{0}
\tag{2.32}\label{eq:2.32}
\end{equation}

\noindent\textbf{Theorem 2.1 (UOP as a single real integral).} Under Assumption 2.1, the value of the UOP at time zero is

\begin{equation}
UOP_{0}=P(0,T)\frac{2}{\pi}\sqrt{KF_{0}}\int_{0}^{\infty} \frac{\sin\left(u(h-x_{0})\right)\sin\left(u(h-k)\right)}{u^{2}+\frac{1}{4}}\Phi_{T}\left(\frac{u^{2}+\frac{1}{4}}{2}\right)du
\tag{2.33}\label{eq:2.33}
\end{equation}

\begin{proof}

The undiscounted value of the up-and-out put value at time 0 is given by

\begin{equation}
KQ^{T}(X_{T}\leq k, M_{T}<h)-E^{\mathbb{Q}^{T}}\left[F_{T}1_{{X_{T}\leq k, M_{T}<h}}\right]
\tag{2.34}\label{eq:2.34}
\end{equation}

Let $Q^{(1)}$ be the measure equivalent to $\mathbb{Q}^{T}$ such that

\begin{equation}
\frac{dQ^{(1)}}{dQ^{T}}\mid_{F_{T}}=\frac{F_{T}}{F_{0}}=e^{X_{T}-x_{0}}.
\tag{2.35}\label{eq:2.35}
\end{equation}

Then

\begin{equation}
E^{\mathbb{Q}^{T}}\left[F_{T}1_{{X_{T}\leq k, M_{T}<h}}\right]=F_{0}Q^{(1)}(X_{T}\leq k, M_{T}<h)
\tag{2.36}\label{eq:2.36}
\end{equation}

Under $Q^{(1)}$, the clock remains independent and the operational drift shifts by $+1$; conditional on $\Gamma_{T}=g$, this is a Cameron--Martin tilt, so

\begin{equation}
X_{T}=x_{0}+(\beta+1)\Gamma_{T}+B_{\Gamma_{T}}^{*} under Q^{(1)}
\tag{2.37}\label{eq:2.37}
\end{equation}

for a Brownian motion $B^{*}$ independent of $\Gamma$. Hence, with $\beta=-\frac{1}{2}$,

\begin{equation}
\mathbb{Q}^{T}(X_{T}\leq k, M_{T}<h)=J_{-1/2}(k;h), Q^{(1)}(X_{T}\leq k, M_{T}<h)=J_{1/2}(k;h)
\tag{2.38}\label{eq:2.38}
\end{equation}

Therefore,

\begin{equation}
UOP_{0}=P(0,T)\left(KJ_{-1/2}(k;h)-F_{0}J_{1/2}(k;h)\right)
\tag{2.39}\label{eq:2.39}
\end{equation}

Insert \eqref{eq:2.30} for $J_{\pm1/2}(k;h)$. Since $\beta^{2}=\frac{1}{4}$ for both signs, the Laplace argument coincides and the combination simplifies algebraically to \eqref{eq:2.33}.

\end{proof}

\subsection{Down-and-out call: single real integral valuation formula by symmetry}

Let the (constant) forward lower barrier be $L>0$, with $L<F_{0}$, and $l:=\log L$

\noindent\textbf{Theorem 2.2 (DOC as a single real integral).} Under Assumption 2.1,

\begin{equation}
DOC_{0}=P(0,T)\frac{2}{\pi}\sqrt{KF_{0}}\int_{0}^{\infty} \frac{\sin\left(u(x_{0}-l)\right)\sin\left(u(k-l)\right)}{u^{2}+\frac{1}{4}}\Phi_{T}\left(\frac{u^{2}+\frac{1}{4}}{2}\right)du
\tag{2.40}\label{eq:2.40}
\end{equation}

\begin{proof}

Let $m_{T}:=inf_{0\leq t\leq T}X_{t}$. The undiscounted down-and-out call value at time 0 is

\begin{equation}
E^{\mathbb{Q}^{T}}\left[F_{T}1_{{X_{T}\geq k, m_{T}>l}}\right]-KQ^{T}(X_{T}\geq k, m_{T}>l)
\tag{2.41}\label{eq:2.41}
\end{equation}

Using \eqref{eq:2.35},

\begin{equation}
E^{\mathbb{Q}^{T}}\left[F_{T}1_{{X_{T}\geq k, m_{T}>l}}\right]=F_{0}Q^{(1)}(X_{T}\geq k, m_{T}>l)
\tag{2.42}\label{eq:2.42}
\end{equation}

Define the reflection about $l$,

\begin{equation}
X^{\sim}_{t}:=2l-X_{t}
\tag{2.43}\label{eq:2.43}
\end{equation}

Then $\{m_{T}>l\}$ is equivalent to $\{\sup_{0\leq t\leq T}X^{\sim}_{t}<l\}$, and $\{X_{T}\geq k\}$ is equivalent to $\{X^{\sim}_{T}\leq 2l-k\}$. From \eqref{eq:2.9},

\begin{equation}
X^{\sim}_{t}=(2l-x_{0})+(-\beta)\Gamma_{t}+B^{\sim}_{\Gamma_{t}}
\tag{2.44}\label{eq:2.44}
\end{equation}

where $B^{\sim}:=-B$ is a Brownian motion independent of $\Gamma$. Under $Q^{(1)}$, the drift of $X$ becomes $\beta+1$ by \eqref{eq:2.37}, hence the drift of $X^{\sim}$ becomes $-(\beta+1)$. Therefore,

\begin{equation}
\mathbb{Q}^{T}\left(X_{T}\geq k, m_{T}>l\right)=J^{\sim}_{-\beta}\left(2l-k;l\right), Q^{\left(1\right)}(X_{T}\geq k, m_{T}>l)=J^{\sim}_{-\left(\beta+1\right)}(2l-k;l)
\tag{2.45}\label{eq:2.45}
\end{equation}

where $J^{\sim}$ is the same function as in \eqref{eq:2.30} but with initial point replaced by $2l-x_{0}$ and upper barrier replaced by $l$.

Specializing to $\beta=-\frac{1}{2}$ yields the same $\pm\frac{1}{2}$ drifts as in the UOP case, hence the same cancellation after inserting \eqref{eq:2.30}: writing $DOC_{0}=P(0,T)\left(F_{0}Q^{(1)}(\cdot)-KQ^{T}(\cdot)\right)$ and using \eqref{eq:2.45}, the integral representation reduces algebraically to \eqref{eq:2.40}.

\end{proof}

\section{Double knock-out options under an independent stochastic clock: discrete Laplace-grid series and explicit coefficients}

This section derives a modular valuation formula for continuously monitored double knock-out (DKO) options in the independent stochastic-clock regime. In contrast with the single-barrier UOP/DOC formulas of Section 2, the natural analytic output for a corridor $(l,h)$ is a real sine series (Dirichlet spectral expansion). The dependence on the volatility specification is entirely captured by the clock Laplace transform $\Phi_{T}$ evaluated on a positive discrete grid, yielding a cacheable pricing engine.

\subsection{Contract and corridor geometry}

Fix constant barriers $L<F_{0}<H$ in the martingale units of Section 2 and a strike $K>0$. Use the log-coordinates $h:=\log H$, $k:=\log K$, $x_{0}:=\log F_{0}$ from \eqref{eq:2.32}, together with $l:=\log L$ from Theorem 2.2, and define the corridor width $a:=h-l>0$.

Consider the exit times $\tau_{l,h}$ and $\sigma_{l,h}$ from the open interval $(l,h)$ as defined as in 2.1.

The value of the DKO at time 0 is given by

\begin{equation}
DKO_{0}=P(0,T)E^{\mathbb{Q}^{T}}\left[\left(e^{X_{T}}-K\right)^{+}1_{\{\tau_{l,h}>T\}}\right]
\tag{3.1}\label{eq:3.1}
\end{equation}

\subsection{Corridor killed kernel and the discrete Laplace grid}

Let

\begin{equation}
\omega_{n}:=\frac{n\pi}{a}, n\geq 1
\tag{3.2}\label{eq:3.2}
\end{equation}

be the Dirichlet eigenfrequencies on $(l,h)$. Define the associated discrete Laplace grid

\begin{equation}
\lambda_{n}:=\frac{1}{2}\left(\omega_{n}^{2}+\beta^{2}\right), n\geq 1
\tag{3.3}\label{eq:3.3}
\end{equation}

Under the martingale normalization $\beta=-\frac{1}{2}$,

\begin{equation}
\lambda_{n}=\frac{1}{2}\left(\omega_{n}^{2}+\frac{1}{4}\right)
\tag{3.4}\label{eq:3.4}
\end{equation}

The defining feature of the independent-clock setting is that the corridor survival/killed law is obtained by mixing the classical Dirichlet heat kernel with respect to the clock $\Gamma_{T}$, and therefore depends on the volatility specification only through the scalar array $(\Phi_{T}(\lambda_{n}))_{n\geq 1}$.

\subsection{Main valuation formula: DKO call as a real sine series}

Let $c:=\max(k,l)$. Define the payoff projection coefficients $(A_{n})_{n\geq 1}$ by

\begin{equation}
A_{n}:=\int_{c}^{h} \left(e^{x}-K\right)e^{\beta x}\sin\left(\omega_{n}(x-l)\right)dx, (n\geq 1)
\tag{3.5}\label{eq:3.5}
\end{equation}

with the convention that $A_{n}=0$ if $c\geq h$ (equivalently, if $k\geq h$). An explicit closed form is obtained by introducing, for $\alpha\in \mathbb{R}$,

\begin{equation}
F_{\alpha,n}(x):=\int e^{\alpha x}\sin\left(\omega_{n}(x-l)\right)dx=\frac{e^{\alpha x}}{\alpha^{2}+\omega_{n}^{2}}\left(\alpha\sin\left(\omega_{n}(x-l)\right)-\omega_{n}\cos\left(\omega_{n}(x-l)\right)\right)
\tag{3.6}\label{eq:3.6}
\end{equation}

so that

\begin{equation}
A_{n}=\left[F_{\beta+1,n}(x)\right]_{x=c}^{x=h}-K\left[F_{\beta,n}(x)\right]_{x=c}^{x=h}
\tag{3.7}\label{eq:3.7}
\end{equation}

Since $\omega_{n}(h-l)=n\pi$, one has $\sin(\omega_{n}(h-l))=0$ and $\cos(\omega_{n}(h-l))=(-1)^{n}$, hence

\begin{equation}
F_{\alpha,n}(h)=(-1)^{n+1}\frac{\omega_{n}e^{\alpha h}}{\alpha^{2}+\omega_{n}^{2}}
\tag{3.8}\label{eq:3.8}
\end{equation}

\noindent\textbf{Theorem 3.1} (DKO call as a discrete Laplace-transform series). Assume $x_{0}\in(l,h)$ and the independent stochastic-clock setting of Section 2. Then

\begin{equation}
DKO_{0}=P(0,T)\frac{2}{a}e^{-\beta x_{0}}\sum_{n=1}^{\infty} \sin\left(\omega_{n}(x_{0}-l)\right)A_{n}\Phi_{T}(\lambda_{n})
\tag{3.9}\label{eq:3.9}
\end{equation}

\begin{proof}

Using independence of $B$ and $\Gamma$, conditioning on $\Gamma_{T}=g$ gives

\begin{equation}
E^{\mathbb{Q}^{T}}\left[\left(e^{X_{T}}-K\right)^{+}1_{\{\tau_{l,h}>T\}} \mid \Gamma_{T}=g\right]=\mathbb{E}\left[\left(e^{Y_{g}}-K\right)^{+}1_{{\sigma_{l,h}>g}}\right]
\tag{3.10}\label{eq:3.10}
\end{equation}

Let $p_{(l,h)}^{(\beta)}(g;x,y)$ be the transition density at time $g$ of $Y$ killed upon exiting $(l,h)$. Then

\begin{equation}
\mathbb{E}\left[\left(e^{Y_{g}}-K\right)^{+}1_{\{\sigma>g\}}\right]=\int_{l}^{h} \left(e^{y}-K\right)^{+}p_{(l,h)}^{(\beta)}(g;x_{0},y)dy
\tag{3.11}\label{eq:3.11}
\end{equation}

Let $q_{(l,h)}(g;x,y)$ denote the driftless Dirichlet heat kernel on $(l,h)$. By formulating the problem in terms of the Kolmogorov equation as in \eqref{eq:2.24}--\eqref{eq:2.25}, one obtains on $(l,h)$

\begin{equation}
p_{(l,h)}^{(\beta)}(g;x,y)=e^{\beta(y-x)-\frac{1}{2}\beta^{2}g}q_{(l,h)}(g;x,y)
\tag{3.12}\label{eq:3.12}
\end{equation}

Moreover, the Dirichlet eigenfunction expansion yields

\begin{equation}
q_{(l,h)}(g;x,y)=\frac{2}{a}\sum_{n=1}^{\infty} e^{-\frac{1}{2}\omega_{n}^{2}g}\sin\left(\omega_{n}(x-l)\right)\sin\left(\omega_{n}(y-l)\right)
\tag{3.13}\label{eq:3.13}
\end{equation}

Inserting \eqref{eq:3.12}--\eqref{eq:3.13} into \eqref{eq:3.11} and using $x=x_{0}$, we obtain for each fixed $g>0$

\begin{equation}
\int_{l}^{h} \left(e^{y}-K\right)^{+}p_{(l,h)}^{(\beta)}(g;x_{0},y)dy=\frac{2}{a}e^{-\beta x_{0}}\sum_{n=1}^{\infty} \sin\left(\omega_{n}(x_{0}-l)\right)A_{n}e^{-\lambda_{n}g}
\tag{3.14}\label{eq:3.14}
\end{equation}

where $A_{n}$ and $\lambda_{n}$ are given by \eqref{eq:3.7} and \eqref{eq:3.4} respectively. Unconditioning with $g=\Gamma_{T}$ yields

\begin{equation}
E^{\mathbb{Q}^{T}}\left[\left(e^{X_{T}}-K\right)^{+}1_{\{\tau_{l,h}>T\}}\right]=E^{\mathbb{Q}^{T}}\left[\frac{2}{a}e^{-\beta x_{0}}\sum_{n=1}^{\infty} \sin\left(\omega_{n}(x_{0}-l)\right)A_{n}e^{-\lambda_{n}\Gamma_{T}}\right]
\tag{3.15}\label{eq:3.15}
\end{equation}

Assume the mild negative-moment condition $E^{\mathbb{Q}^{T}}\left[\Gamma_{T}^{-1/2}<\infty\right]$ already used in Section 2.

First, one has the bound

\begin{equation}
\mid A_{n}\mid\leq\frac{C_{1}}{\omega_{n}}, n\geq 1
\tag{3.16}\label{eq:3.16}
\end{equation}

for some constant $C_{1}$ depending only on $(l,h,k,K,\beta)$. Indeed, $\left(e^{x}-K\right)^{+}e^{\beta x}$ is bounded and piecewise $C^{1}$ on $[l,h]$ (with the only kink at $x=k$); integrating by parts once on each smooth subinterval yields the stated $O(\omega_{n}^{-1})$ decay.

Second, for $\lambda>0$ and $z>0$, the inequality $e^{-\lambda z}\leq\Gamma(3/2)(\lambda z)^{-1/2}$ gives

\begin{equation}
\Phi_{T}(\lambda)=E^{\mathbb{Q}^{T}}\left[e^{-\lambda\Gamma_{T}}\right]\leq\Gamma(3/2)\lambda^{-1/2}E^{\mathbb{Q}^{T}}\left[\Gamma_{T}^{-1/2}\right]
\tag{3.17}\label{eq:3.17}
\end{equation}

Since $\lambda_{n}\sim\frac{1}{2}\omega_{n}^{2}$, this implies $\Phi_{T}(\lambda_{n})\leq C_{2}/\omega_{n}$ for some constant $C_{2}$. Therefore,

\begin{equation}
\sum_{n=1}^{\infty} \mid\sin\left(\omega_{n}(x_{0}-l)\right)A_{n}\Phi_{T}(\lambda_{n})\mid\leq\sum_{n=1}^{\infty} \frac{C_{1}C_{2}}{\omega_{n}^{2}}<\infty
\tag{3.18}\label{eq:3.18}
\end{equation}

so that the interchange of expectation with the infinite series is permissible in \eqref{eq:3.15}, giving \eqref{eq:3.9} after multiplication by $P(0,T)$.

\end{proof}

\subsection{Remarks and discussion}

(i) The series \eqref{eq:3.9} is purely real and requires only elementary sine/exponential evaluations and $\Phi_{T}$ on the real grid $\{\lambda_{n}\}$. No complex integration is needed. The clock enters only through the evaluations

\begin{equation}
\Phi_{T}(\lambda_{n})=E^{\mathbb{Q}^{T}}\left[e^{-\lambda_{n}\Gamma_{T}}\right], \lambda_{n}=\frac{1}{2}(\omega_{n}^{2}+\beta^{2}), n\geq 1
\tag{3.19}\label{eq:3.19}
\end{equation}

For fixed $(l,h,T)$, the grid $\{\lambda_{n}\}$ is fixed. Hence $(\Phi_{T}(\lambda_{n}))_{n\leq N}$ can be precomputed once and reused across strikes $K$, payoffs, and even across repeated calibration iterations where only the clock parameters change.

(ii) Spectral expansions for double-barrier options under independent time changes are a known theme in the literature (see, e.g. (Hieber and Scherer, 2012)). The contribution here is to integrate the corridor series into the same transform-only architecture as the single-barrier one-integral formulas of Section 2, and to position it as the baseline $\rho=0$ layer for the leverage corrections developed later.

(iii) DKO put. For a DKO put payoff $\left(K-e^{X_{T}}\right)^{+}1_{\{\tau_{l,h}>T\}}$, the same series \eqref{eq:3.9} holds with coefficients

\begin{equation}
A_{n}^{put}:=\int_{l}^{d} \left(K-e^{x}\right)e^{\beta x}\sin\left(\omega_{n}(x-l)\right)dx, d:=\min(k,h)
\tag{3.20}\label{eq:3.20}
\end{equation}

(with $A_{n}^{put}=0$ if $d\leq l$), and the same $\omega_{n},\lambda_{n}$ as in \eqref{eq:3.2}--\eqref{eq:3.4}

\section{Solvable stochastic clocks: explicit and low-dimensional Laplace transforms}

This section collects classes of stochastic clocks for which the transform of the terminal clock is available in closed form or via low-dimensional deterministic systems. In all cases, the formulas of Sections 2--3 remain plug-in: once the clock transform is evaluable on the required real arguments, pricing reduces to one-dimensional real quadratures (single barriers) or rapidly convergent real series (double barriers).

For single-barrier formulas (Section 2), $\Phi_{T}$ is evaluated along the one-dimensional quadrature line $\lambda=\frac{u^{2}+\beta^{2}}{2},u\in[0,\infty),$ with $\beta=-\frac{1}{2}$ under the forward martingale normalization. For double-barrier series (Section 3), $\Phi_{T}$ is evaluated on the discrete grid $(\omega_{n},\lambda_{n})$ given by \eqref{eq:3.2}--\eqref{eq:3.4}.

\subsection{Integrated CIR clock and time-inhomogeneous affine extensions}

A canonical continuous clock is obtained by taking the activity $v$ to be a square-root (CIR) diffusion, independent of the Brownian motion driving the log-forward (Assumption 2.1). This yields a time-changed Brownian analogue of a Heston-type activity factor, with strict positivity and mean reversion.

Let $(v_{t})_{t\leq T}$ solve, under $\mathbb{Q}^{T}$,

\begin{equation}
dv_{t}=\kappa(\theta-v_{t})dt+\xi\sqrt{v_{t}}dZ_{t}, v_{0}>0
\tag{4.1}\label{eq:4.1}
\end{equation}

where $Z$ is a Brownian motion independent of the Brownian motion in the log-forward representation \eqref{eq:2.9}. For each $\lambda\geq 0$,

\begin{equation}
\Phi_{T}(\lambda)=\exp\left(-A(T;\lambda)-B(T;\lambda)v_{0}\right)
\tag{4.2}\label{eq:4.2}
\end{equation}

where $B$ solves the Riccati ODE

\begin{equation}
\frac{d}{dt}B(t;\lambda)=\kappa B(t;\lambda)+\lambda-\frac{1}{2}\xi^{2}B(t;\lambda)^{2}, B(0;\lambda)=0,
\tag{4.3}\label{eq:4.3}
\end{equation}

and $A$ is obtained by quadrature,

\begin{equation}
\frac{d}{dt}A(t;\lambda)=\kappa\theta B(t;\lambda), A(0;\lambda)=0
\tag{4.4}\label{eq:4.4}
\end{equation}

For constant $(\kappa,\theta,\xi)$, a classical exact solution is easily obtained (Cox et al., 1985).

The integrated CIR clock provides (i) strict positivity of the activity $v$, (ii) parsimonious mean reversion with a tight, interpretable parameter set, and (iii) fast calibration through explicit transforms due to maximum tractability per parameter.

A limitation is that a single CIR factor imposes essentially one persistence scale for volatility clustering. Thus, it may struggle to reproduce the empirically observed multi-horizon decay of volatility autocorrelations unless one allows additional structure. A second limitation is calibration identifiability: the triplet $(\kappa,\theta,\eta)$ can trade off in their impact on $\Gamma_{T}$, especially when one calibrates primarily to European smiles at a small number of maturities. In other words, from the option data you're fitting, there may not be enough information to uniquely pin down $(\kappa,\theta,\eta)$. Different parameter triples can produce almost the same distribution of the integrated variance and therefore almost the same European smile.

To accommodate an activity term structure while preserving tractability, one may allow piecewise-continuous time-dependent coefficients $\kappa(t),\theta(t),\xi(t)$. The transform remains exponential-affine as in \eqref{eq:4.2}, with $B$ solving the time-dependent Riccati ODE

\begin{equation}
\frac{d}{dt}B(t;\lambda)=\kappa(t)B(t;\lambda)+\lambda-\frac{1}{2}\xi(t)^{2}B(t;\lambda)^{2}, B(0;\lambda)=0
\tag{4.5}\label{eq:4.5}
\end{equation}

and

\begin{equation}
A(T;\lambda)=\int_{0}^{T} \kappa(t)\theta(t)B(t;\lambda)dt
\tag{4.6}\label{eq:4.6}
\end{equation}

The computational overhead is one scalar ODE per $\lambda$, negligible compared to the barrier layer, while the benefit is a better fit of both short- and long-run variance dynamics

\subsection{Squared OU clock and quadratic--Gaussian generalizations}

Another continuous clock family is obtained from Gaussian factors with quadratic activity. This retains analytic Laplace transforms while naturally producing multiple time scales and smooth volatility clustering.

Let $Y$ be an Ornstein--Uhlenbeck factor

\begin{equation}
dY_{t}=-\alpha Y_{t}dt+\sigma\, dZ_{t}, Y_{0}\in \mathbb{R}
\tag{4.7}\label{eq:4.7}
\end{equation}

with $Z$ independent of the Brownian motion in \eqref{eq:2.9}, and define the nonnegative activity and clock

\begin{equation}
v_{t}:=Y_{t}^{2}, \Gamma_{T}=\int_{0}^{T} Y_{s}^{2}ds
\tag{4.8}\label{eq:4.8}
\end{equation}

Then the transform has the exponential--quadratic form

\begin{equation}
\Phi_{T}(\lambda)=\exp\left(-A(T;\lambda)-B(T;\lambda)Y_{0}^{2}\right)
\tag{4.9}\label{eq:4.9}
\end{equation}

where $B$ solves the scalar Riccati ODE

\begin{equation}
\frac{d}{dt}B(t;\lambda)=2\sigma^{2}B(t;\lambda)^{2}-2\alpha B(t;\lambda)+\lambda, B(0;\lambda)=0
\tag{4.10}\label{eq:4.10}
\end{equation}

and $A$ is obtained by quadrature:

\begin{equation}
\frac{d}{dt}A(t;\lambda)=\sigma^{2}B(t;\lambda), A(0;\lambda)=0
\tag{4.11}\label{eq:4.11}
\end{equation}

The exact solution is classical; see Stein and Stein (1991) and Dankel (1991) for seminal references.

This clock is attractive when one wants Gaussian latent factors (easy simulation/filtering) and the ability to superpose several OU modes to fit empirical volatility autocorrelation. It can be extended to a multi-factor version. Let $Y_{t}\in \mathbb{R}^{d}$ be a Gaussian Markov factor (e.g. a multi-factor OU)

\begin{equation}
dY_{t}=K(\mu-Y_{t})dt+\Sigma\, dZ_{t}
\tag{4.12}\label{eq:4.12}
\end{equation}

with $Z$ independent of the Brownian motion in \eqref{eq:2.9}, and let the activity be a nonnegative quadratic form

\begin{equation}
v_{t}:=Y_{t}^{\top}QY_{t}+q^{\top}Y_{t}+q_{0},\qquad Q\succeq 0,\quad v_{t}\geq 0
\tag{4.13}\label{eq:4.13}
\end{equation}

Then the transform takes the exponential--quadratic form

\begin{equation}
\Phi_{T}(\lambda)=\exp\left(-A(T;\lambda)-b(T;\lambda)^{\top}Y_{0}-Y_{0}^{\top}C(T;\lambda)Y_{0}\right)
\tag{4.14}\label{eq:4.14}
\end{equation}

where $C$ solves a matrix Riccati ODE, $b$ a linear ODE driven by $C$, and $A$ is obtained by quadrature. These systems are standard in quadratic--Gaussian transform theory and can be solved robustly with off-the-shelf ODE solvers.

Beyond easy simulation/filtering, the key advantage of the squared OU clock is that it provides a smooth and naturally mean-reverting activity mechanism while allowing multiple time scales through superposition of OU modes. In the independent-clock barrier setting, this is valuable because the maturity-dependence of barrier prices is largely controlled by how the distribution of $\Gamma_{T}$ evolves with $T$: adding OU modes is a direct way to enrich that evolution without sacrificing analytic (ODE-based) Laplace evaluation.

A limitation is that quadratic-Gaussian activity remains ``diffusion-like'': it can generate realistic clustering, but may be less effective at producing sharp volatility bursts unless parameters are pushed into regimes that can distort other features (e.g., too much mass of $v_{t}$ near $0$ because $X_{t}$ frequently visits the origin). From a numerical standpoint, the one-factor squared OU case is very stable, but the fully general quadratic--Gaussian extension requires solving matrix Riccati systems; while standard, these systems can become stiff for large factor dimension, long maturities, or large Laplace arguments $q$, precisely the regime that may arise when $\Phi_{T}$ is sampled over a wide range of $u$ in the barrier integrals.

Accordingly, the natural extension route is to keep dimension modest. A small multi-factor OU (two or three modes) often captures most of the empirical autocorrelation structure, whereas high-dimensional quadratic forms can create calibration non-uniqueness: different $(Q,\kappa,\Sigma)$ combinations can yield similar $\Gamma_{T}$ marginals.

\subsection{Markov-switching and phase-type clocks}

Barrier pricing is sensitive to persistent volatility regimes (calm vs stressed markets). Markov-switching clocks introduce such regimes without introducing jumps in the log-price itself: the asset path remains continuous because only the activity rate changes.

Let $(R_{t})_{t\leq T}$ be a continuous-time Markov chain on $\{1,\cdots,m\}$ with generator matrix $Q$, independent of the Brownian motion in \eqref{eq:2.9}. Fix regime activity levels $v_{1},\cdots,v_{m}\geq 0$ and set

\begin{equation}
v_{t}:=v_{R_{t}}, \Gamma_{T}=\int_{0}^{T} v_{R_{s}}ds
\tag{4.15}\label{eq:4.15}
\end{equation}

Let $\alpha\in \mathbb{R}^{m}$ be the initial distribution of $R_{0}$, let $D:=\operatorname{diag}(v_{1},\cdots,v_{m})$, and let $1\in \mathbb{R}^{m}$ be the vector of ones. Then, for $\lambda\geq 0$,

\begin{equation}
\Phi_{T}(\lambda)=\alpha^{\top}\exp\left((Q-\lambda D)T\right)\mathbf{1}
\tag{4.16}\label{eq:4.16}
\end{equation}

This follows from a backward Kolmogorov argument and reduces transform evaluation to deterministic linear algebra.

Markov-switching clocks are appealing for barrier products because they embed persistent regime behaviour directly: extended periods of high activity (stress) materially increase barrier crossing probabilities, and a finite-state chain gives an explicit mechanism to represent such persistence. Computationally, the transform evaluation reduces to deterministic linear algebra (matrix exponentials), which is well-suited to the repeated evaluations required by Sections 2--3.

The main modeling limitation is the stylized nature of piecewise-constant activity: volatility changes instantaneously when the regime switches. This can be realistic at coarser time scales, but it may be too rough for high-frequency interpretations, and it can produce a ``blocky'' dynamics for $\Gamma_{t}$ unless the state space is refined. Increasing the number of regimes improves realism but introduces a state-dimension vs calibration stability trade-off: many-state chains can be hard to identify from option data, and repeated matrix exponentials become more costly (though still typically cheap relative to the barrier layer).

Phase-type (PH) specifications push this idea further by offering a flexible matrix-exponential class that can approximate a wide range of nonnegative laws for $\Gamma_{T}$. The advantage is shape control (including multi-modality and flexible tail behaviour) while retaining the same transform interface. The limitation is that PH representations are often non-unique: many distinct sub-generators can fit the same marginal distribution, and without structural constraints (e.g., Coxian/Erlang mixtures) calibration can become ill-conditioned. Moreover, if PH is used purely as a marginal fit at each maturity, one should be cautious about dynamic consistency across $T$: matching $\Gamma_{T}$ at each $T$ does not automatically define a coherent clock process unless the PH structure is imposed at the process level.

\subsection{Affine clocks with jumps in the variance driver}

To capture sudden spikes in volatility activity (news shocks, liquidity events) while maintaining continuous asset paths, one may introduce jumps in the activity factor $v$ but keep the log-price driven by Brownian motion time-changed by the integrated activity. The barrier methodology of Sections 2--3 continues to apply because $\Gamma$ remains continuous whenever $v$ is càdlàg and integrable.

Let $v$ be a one-dimensional affine jump-diffusion (or, more generally, an affine process) on $R_{+}$, independent of the Brownian motion in \eqref{eq:2.9}, and define $\Gamma_{T}=\int_{0}^{T} v_{s}ds$. Affine transform theory yields an exponential-affine Laplace transform of the form

\begin{equation}
\Phi_{T}(\lambda)=\exp\left(-A(T;\lambda)-B(T;\lambda)v_{0}\right)
\tag{4.17}\label{eq:4.17}
\end{equation}

where $B$ solves a generalized Riccati equation featuring an additional term characterizing the jump compensator (via the Lévy exponent), and $A$ is obtained by quadrature as in the purely diffusive affine case. The precise form is model-dependent but standard.

From a modeling standpoint, jump-activity clocks offer two advantages: (i) they generate realistic short-lived volatility bursts that are difficult to reproduce with purely diffusive factors, and (ii) they can improve calibration to short-maturity smiles while preserving the low-dimensional transform interface required by Sections 2--3.

Jump-activity clocks extend the diffusive affine family in a way that is particularly relevant for barrier pricing: by allowing jumps in $v_{t}$ while keeping the log-price time-changed Brownian, one can model abrupt increases in activity (news, liquidity events) without introducing discontinuities in the asset path. This improves the realism of short-horizon barrier risk (where sudden surges in realized variance matter) while preserving the same low-dimensional transform workflow: $\Phi_{T}(q)$ is still obtained from generalized Riccati systems, evaluated at real $q\geq 0$.

The main limitation is parameter proliferation and identifiability. Jump intensity and jump-size parameters can substitute for diffusive vol-of-vol in their effect on $\Gamma_{T}$, so calibration may require strong regularization or joint fitting across maturities.

In addition, we stress a fundamental scope limitation: if one time-changes Brownian motion by a pure-jump subordinator, the asset typically develops jumps. Barrier pricing then requires different techniques (nonlocal generators / Wiener--Hopf) and lies outside the continuous-path framework adopted here.

\section{Leverage as a correlation perturbation around the independent-clock barrier engine}

Sections 2--3 provide simple, explicit single- and double-barrier valuation formulas in the independent-clock regime, i.e. when the Brownian driver of returns is independent of the factor(s) driving the stochastic clock. In equity and FX markets, however, barrier prices are strongly affected by leverage, understood here as nonzero instantaneous correlation between return shocks and volatility/clock shocks. When such correlation is present, the one-dimensional time-change reduction breaks down and barrier pricing becomes a genuinely two-factor killed boundary problem with mixed derivatives.

This section introduces a modular leverage layer by treating correlation $\rho$ as a perturbation parameter and expanding barrier values around the solvable anchor $\rho=0$. The key structural points are:

(i) Correlation enters the two-factor generator linearly through a single mixed-derivative operator.

(ii) The resulting $\rho$-expansion yields a forced hierarchy under the same ``decoupled'' killed operator $\mathcal{L}_{0}$ that governs the baseline.

(iii) Duhamel's principle (Dyson--Phillips expansion) yields a semigroup representation of each coefficient as a time integral of expectations under the $\rho=0$ killed dynamics, enabling efficient numerical routes.

Throughout, we keep the development clock-modular: model specificity enters through the conditional Laplace transforms described in Section 4.

\subsection{Correlated clock-driven diffusion under the $T$-forward measure}

We work under the $T$-forward measure $\mathbb{Q}^{T}$ and in the martingale normalization of Sections 2--3 (discounted/forward underlying). Let $X_{t}=\log F_{t}$ be the log-forward (or log-discounted) price, and let $Y_{t}$ be a Markov factor driving the instantaneous activity rate

\begin{equation}
v(Y_{t})>0
\tag{5.1}\label{eq:5.1}
\end{equation}

We consider the two-factor diffusion

\begin{equation}
dX_{t}=\beta v(Y_{t})dt+\sqrt{v(Y_{t})}\,dW_{t}^{(1)},\qquad dY_{t}=b(Y_{t})dt+a(Y_{t})\,dW_{t}^{(2)}
\tag{5.2}\label{eq:5.2}
\end{equation}

where $(W^{(1)},W^{(2)})$ is a two-dimensional Brownian motion under $\mathbb{Q}^{T}$ with constant instantaneous correlation

\begin{equation}
\langle W^{(1)},W^{(2)}\rangle _{t}=\rho\,dt, \rho\in[-1,1]
\tag{5.3}\label{eq:5.3}
\end{equation}

and we assume the SDE in \eqref{eq:5.2} is well posed in the weak sense: for each $y\in E$ there exists a non-explosive weak solution $Y$ with $Y_{0}=y$, and uniqueness in law holds (equivalently, the martingale problem for the $Y$-generator is well posed).

Define the stochastic clock

\begin{equation}
\Gamma_{t}:=\int_{0}^{t} v(Y_{s})ds
\tag{5.4}\label{eq:5.4}
\end{equation}

\noindent\textbf{Remark 5.1} : In affine one-factor cases one may take $Y\equiv v$ and specify the model as follows

\begin{equation}
dX_{t}=\beta v(Y_{t})dt+\sqrt{v(Y_{t})}\,dW_{t}^{(1)},\qquad dY_{t}=b(Y_{t})dt+a(Y_{t})\,dW_{t}^{(2)}
\tag{5.5}\label{eq:5.5}
\end{equation}

We introduce $Y$ to cover the broader clock families of Section 4 (quadratic/multi-factor/switching), where the transform is naturally conditional on the factor state rather than on the instantaneous activity alone.

\subsection{Barrier value functions and killed boundary problems}

Let $D\subset \mathbb{R}\times E$ be the admissible state region (in log-space), where $E$ is the state space of $Y$. The two principal geometries used in this paper are the single upper barrier (in log-space) $D=(-\infty,h)\times E$ and the corridor (double barrier) $D=(l,u)\times E$. Let $\tau$ be the first exit time of $(X,Y)$ from $D$. For a measurable payoff $\Psi:\mathbb{R}\to \mathbb{R}$ (e.g. $\Psi(x)=(e^{x}-K)^{+}$ in forward units), define the time-$t$ barrier value function under correlation $\rho$ by

\begin{equation}
u^{\rho}(t,x,y):=E_{t,x,y}^{\mathbb{Q}^{T}}\left[\Psi(X_{T})1_{\{\tau>T\}}\right]
\tag{5.6}\label{eq:5.6}
\end{equation}

Discounting is absorbed by the $T$-forward numeraire. The time-$t$ price in domestic units is $V_{t}=P\left(t,T\right) u^{\rho}\left(t,X_{t},Y_{t}\right)$; cf. Section 2.

Under standard regularity assumptions (classical solution) or in the mild/viscosity sense, $u^{\rho}$ solves the killed backward problem on $D$:

\begin{equation}
\left\{\begin{aligned}(\partial_{t}+\mathcal{L}_{\rho})u^{\rho}(t,x,y)&=0, &&(t,x,y)\in[0,T)\times D,\\u^{\rho}(t,x,y)&=0, &&(t,x,y)\in[0,T)\times\partial D,\\u^{\rho}(T,x,y)&=\Psi(x), &&(x,y)\in D.\end{aligned}\right.
\tag{5.7}\label{eq:5.7}
\end{equation}

\subsection{Generator splitting: $\mathcal{L}_{\rho}=\mathcal{L}_{0}+\rho \mathcal{L}_{1}$}

Write partial derivatives as subscripts. The generator associated with \eqref{eq:5.2}--\eqref{eq:5.3} is

\begin{equation}
\mathcal{L}_{\rho}f=\beta v(y)f_{x}+b(y)f_{y}+\frac{1}{2}v(y)f_{xx}+\frac{1}{2}a(y)^{2}f_{yy}+\rho a(y)\sqrt{v(y)}f_{xy}
\tag{5.8}\label{eq:5.8}
\end{equation}

Define the decoupled operator (the $\rho=0$ generator) and the correlation perturbation operator by

\begin{equation}
\mathcal{L}_{0}f:=\beta v(y)f_{x}+b(y)f_{y}+\frac{1}{2}v(y)f_{xx}+\frac{1}{2}a(y)^{2}f_{yy}, \mathcal{L}_{1}f:=a(y)\sqrt{v(y)}f_{xy}
\tag{5.9}\label{eq:5.9}
\end{equation}

Then

\begin{equation}
\mathcal{L}_{\rho}=\mathcal{L}_{0}+\rho \mathcal{L}_{1}
\tag{5.10}\label{eq:5.10}
\end{equation}

$\mathcal{L}_{0}$ governs the independent-clock dynamics and is the operator for which Sections 2--3 provide analytical barrier engines (via conditional Laplace transforms). All leverage effects are isolated in the mixed derivative $f_{xy}$, scaled by $\rho$ and the local factor $a(y)\sqrt{v(y)}$.

\subsection{The $\rho$-expansion and forced hierarchy under the killed $\mathcal{L}_{0}$-operator}

We expand $u^{\rho}$ as a formal power series around $\rho=0$:

\begin{equation}
u^{\rho}(t,x,y)=\sum_{n=0}^{\infty} \rho^{n}u_{n}(t,x,y),\qquad u_{n}:=\frac{1}{n!}\partial_{\rho}^{n}u^{\rho}\big|_{\rho=0}
\tag{5.11}\label{eq:5.11}
\end{equation}

Inserting \eqref{eq:5.11} into \eqref{eq:5.7} and using \eqref{eq:5.10}, identification of powers of $\rho$ yields the hierarchy:

(i) Order $0$ (baseline):

\begin{equation}
\left\{\begin{aligned}(\partial_{t}+\mathcal{L}_{0})u_{0}&=0, &&\text{in } [0,T)\times D,\\u_{0}&=0, &&\text{on } [0,T)\times\partial D,\\u_{0}(T,\cdot)&=\Psi(\cdot), &&\text{on } D.\end{aligned}\right.
\tag{5.12}\label{eq:5.12}
\end{equation}

(ii) Order $n\geq 1$ (forced problems):

\begin{equation}
\left\{\begin{aligned}(\partial_{t}+\mathcal{L}_{0})u_{n}&=-\mathcal{L}_{1}u_{n-1}, &&\text{in } [0,T)\times D,\\u_{n}&=0, &&\text{on } [0,T)\times\partial D,\\u_{n}(T,\cdot)&=0, &&\text{on } D.\end{aligned}\right.
\tag{5.13}\label{eq:5.13}
\end{equation}

Each coefficient $u_{n}$ solves a killed boundary problem under the same decoupled operator $\mathcal{L}_{0}$. Correlation enters only through the forcing term $-\mathcal{L}_{1}u_{n-1}$. In particular, one avoids solving a new mixed-derivative two-dimensional PDE for each $\rho$; instead, one reuses the baseline machinery and adds leverage through a forced sequence.

\subsection{Duhamel semigroup representation of the coefficients}

Let $(X^{(0)},Y^{(0)})$ denote the process under $\rho=0$ (i.e. with independent Brownian drivers), and let $\tau^{(0)}$ be its first exit time from $D$. Define the killed $\mathcal{L}_{0}$-semigroup acting on suitable test functions $\phi$ by

\begin{equation}
(P_{t,s}^{(0)}\phi)(x,y):=E_{t,x,y}^{\mathbb{Q}^{T}}\left[\phi\left(X_{s}^{(0)},Y_{s}^{(0)}\right)1_{\{\tau^{(0)}>s\}}\right], 0\leq t\leq s\leq T
\tag{5.14}\label{eq:5.14}
\end{equation}

\noindent\textbf{Proposition 5.1} (Duhamel representation of $u_{n}$). Assume $u_{0}$ solves \eqref{eq:5.12} and, for $n\geq 1$, $u_{n}$ solves \eqref{eq:5.13} in the mild sense. Then for $0\leq t\leq T$,

\begin{equation}
u_{0}(t,\cdot)=P_{t,T}^{(0)}\Psi
\tag{5.15}\label{eq:5.15}
\end{equation}

and for $n\geq 1$,

\begin{equation}
u_{n}(t,\cdot)=-\int_{t}^{T} P_{t,s}^{(0)}\left(\mathcal{L}_{1}u_{n-1}(s,\cdot)\right)ds
\tag{5.16}\label{eq:5.16}
\end{equation}

Equivalently, in expectation form,

\begin{equation}
u_{n}(t,x,y)=-E_{t,x,y}^{\mathbb{Q}^{T}}\left[\int_{t}^{T} \left(\mathcal{L}_{1}u_{n-1}\right)\left(s,X_{s}^{(0)},Y_{s}^{(0)}\right)1_{\{\tau^{(0)}>s\}}ds\right]
\tag{5.17}\label{eq:5.17}
\end{equation}

\begin{proof}

The identity \eqref{eq:5.15} is the standard Feynman--Kac representation for killed diffusions. For $n\geq 1$, apply Duhamel's principle to the inhomogeneous killed equation \eqref{eq:5.13} with terminal condition $u_{n}(T,\cdot)=0$, using the killed evolution family generated by $\mathcal{L}_{0}$. This yields \eqref{eq:5.16}, and \eqref{eq:5.17} follows by the definition of the semigroup \eqref{eq:5.14}.

\end{proof}

\subsection{Baseline evaluator $u_{0}$ as a conditional plug-in of Sections 2--3}

The leverage hierarchy \eqref{eq:5.12}--\eqref{eq:5.13} requires the baseline solution $u_{0}(t,x,y)$ for many intermediate times $t\in[0,T]$, not only at $t=0$. Under $\rho=0$, conditional on the clock-factor state $Y_{t}=y$, the remaining clock increment is

\begin{equation}
\Gamma_{t,T}:=\Gamma_{T}-\Gamma_{t}=\int_{t}^{T} v(Y_{s})ds
\tag{5.18}\label{eq:5.18}
\end{equation}

Denote its state-dependent Laplace transform by

\begin{equation}
\Phi_{t,T}(\lambda;y):=E^{\mathbb{Q}^{T}}\left[e^{-\lambda\Gamma_{t,T}}\mid Y_{t}=y\right], \lambda\geq 0
\tag{5.19}\label{eq:5.19}
\end{equation}

For the affine/quadratic clock families emphasized in Section 4, $\Phi_{t,T}(\lambda;y)$ is available in exponential--affine (or exponential--quadratic) form, typically via Riccati systems, and can be evaluated efficiently on the same $\lambda$-grids used for the unconditional transform $\Phi_{T}$; see Section 4 and standard transform technology.

Conditional on the clock path (equivalently on $\Gamma_{t,T}$), the log-forward evolves as a drifted Brownian motion run at operational time $\Gamma_{t,T}$. Therefore the single- and double-barrier solvers of Sections 2--3 transfer verbatim once $\Phi_{T}(\lambda)$ is replaced by $\Phi_{t,T}(\lambda;y)$. This is the main modularity point of the leverage layer: no new barrier kernel is required---only conditional transform input to the existing baseline engines.

\subsection{First-order forcing and semi-analytic differentiation}

From the definition of $\mathcal{L}_{1}$ in \eqref{eq:5.9}, the forcing operator acting on a sufficiently smooth function $f$ is

\begin{equation}
\mathcal{L}_{1}f(x,y)=a(y)\sqrt{v(y)}\partial_{xy}f(x,y)
\tag{5.20}\label{eq:5.20}
\end{equation}

Hence, in the forced hierarchy \eqref{eq:5.13}, the dominant computational task is the evaluation of mixed derivatives $\partial_{xy}u_{n-1}$ (starting with $u_{0}$) on the interior of the killed domain.

In the independent-clock baseline, $u_{0}(t,x,y)$ is represented either as a one-dimensional real integral (single barriers), or a real sine series (double barriers), where the entire dependence on the clock state $y$ enters only through the conditional transform $\Phi_{t,T}(\cdot;y)$ as in \eqref{eq:5.19}. Consequently:

(i) $\partial_{x}$ acts only on elementary exponential/sine/cosine factors in the barrier kernel;

(ii) $\partial_{y}$ acts only on $\Phi_{t,T}(\lambda;y)$ (and therefore on the Riccati coefficients in affine/quadratic cases).

Under mild dominated convergence conditions (or termwise differentiation for the sine series under standard summability assumptions), one may differentiate under the integral / termwise in the series to obtain $\partial_{xy}u_{0}$ without finite differencing near the barrier. This ``differentiate-the-transform'' mechanism is what keeps the forcing term stable and compatible with calibration loops.

In terms of regularity, if kinks at the strike or the absorbing boundary reduce classical regularity, one may smooth the payoff slightly (or work with weak derivatives). In practice, the semi-analytic differentiation remains well behaved when implemented at the level of the transform input, with a conventional smoothing/weak formulation stated in the numerical section.

\subsection{Practical low-order approximations and stabilization}

In applications one typically truncates the expansion \eqref{eq:5.11} at low order:

\begin{equation}
u^{\rho}(t,x,y)\approx u^{(N)}(t,x,y):=\sum_{n=0}^{N} \rho^{n}u_{n}(t,x,y)
\tag{5.21}\label{eq:5.21}
\end{equation}

The coefficients $u_{n}$ are defined by the forced hierarchy \eqref{eq:5.13} and admit the Duhamel representations \eqref{eq:5.16}--\eqref{eq:5.17}. This supports two complementary numerical routes developed later: (i) forced PDE solves under $\mathcal{L}_{0}$, and (ii) a Duhamel--Monte Carlo estimator under the $\rho=0$ killed dynamics---both avoiding mixed derivatives in the solved dynamics.

For equity-like correlations (often $|\rho|\gtrsim 0.5$), a raw low-order Taylor truncation can be materially improved by minimal resummation. A convenient choice is the $[1/1]$ Padé approximant built from $u_{0},u_{1},u_{2}$:

\begin{equation}
u^{\rho}(t,x,y)\approx \mathcal{P}_{[1/1]}(\rho):=\frac{u_{0}+\rho\left(u_{1}-\frac{u_{0}u_{2}}{u_{1}}\right)}{1-\rho\frac{u_{2}}{u_{1}}},\qquad (u_{1}\neq 0)
\tag{5.22}\label{eq:5.22}
\end{equation}

This introduces a potential pole at $\rho_{\star}=u_{1}/u_{2}$. In practice one monitors pole proximity and falls back to the Taylor truncation \eqref{eq:5.21} when the pole is too close to the target $\rho$. One also enforces basic structural constraints (e.g. nonnegativity of option values and, when pricing survival probabilities, the bound $0\leq \mathrm{survival}\leq 1$).

\subsection{Minimal error diagnostics (continuous level)}

A leverage expansion is useful only if it comes with a low-cost way to detect when it is being pushed beyond its reliable regime. Define the $N$-th order truncation $u^{(N)}$ by \eqref{eq:5.21} and consider its killed-PDE residual in the domain:

\begin{equation}
R_{\rho}^{(N)}(t,x,y):=\left(\partial_{t}+\mathcal{L}_{\rho}\right)u^{(N)}(t,x,y), (t,x,y)\in[0,T)\times D
\tag{5.23}\label{eq:5.23}
\end{equation}

By construction of the hierarchy \eqref{eq:5.12}--\eqref{eq:5.13}, all terms up to order $\rho^{N}$ cancel, and one obtains the explicit leading-order identity

\begin{equation}
R_{\rho}^{(N)}(t,x,y)=\rho^{N+1}\mathcal{L}_{1}u_{N}(t,x,y)
\tag{5.24}\label{eq:5.24}
\end{equation}

This is computable directly from the last retained coefficient $u_{N}$, since $\mathcal{L}_{1}$ is the mixed-derivative operator \eqref{eq:5.20}.

To turn \eqref{eq:5.24} into an a posteriori indicator, use the fact that the killed semigroup associated with a uniformly parabolic diffusion on a bounded truncated domain is a contraction in the sup norm (after numerical truncation of the spatial domain, which is standard in implementation). Then, since the true solution satisfies $(\partial_{t}+\mathcal{L}_{\rho})u^{\rho}=0$ with matching terminal and boundary data, the truncation error $e^{(N)}:=u^{\rho}-u^{(N)}$ solves a killed inhomogeneous problem driven by $-R_{\rho}^{(N)}$. Duhamel's principle yields the practical bound/indicator

\begin{equation}
\|e^{(N)}(t,\cdot)\| _{\infty}\lesssim \int_{t}^{T} \|R_{\rho}^{(N)}(s,\cdot)\| _{\infty}ds=|\rho|^{N+1}\int_{t}^{T} \|\mathcal{L}_{1}u_{N}(s,\cdot)\| _{\infty}ds
\tag{5.25}\label{eq:5.25}
\end{equation}

which can be evaluated on the same time--space grid used to compute $u_{N}$ (using semi-analytic differentiation when available, or finite differences for $\partial_{xy}u_{N}$ if needed).

\section{Numerical Results}

This section measures the performance of the pricing methodology developed in Sections 2--5 on representative single-barrier knock-out contracts, under two stochastic-clock specifications: the integrated CIR clock (Section 4.3) and the squared Ornstein--Uhlenbeck clock (Section 4.4). We proceed in two steps: first, we validate the independent-clock barrier formulas ($\rho=0$) against high-precision Monte Carlo; then, we assess the leverage correction introduced in Section 5 via the $\rho$-expansion, including its stabilization by Padé acceleration for large $|\rho|$.

\begin{center}

\textbf{Table 6.1 -- Fixed contract and market parameters used in the numerical experiments}

\medskip

{\small

\begin{tabular}{l c c}

\hline

\textbf{Parameter} & \textbf{Symbol} & \textbf{Value} \\

\hline

Spot & $S_{0}$ & 100 \\

Strike & $K$ & 100 (ATM) \\

Risk-free rate & $r$ & 0.03 \\

Dividend yield & $q$ & 0 \\

Lower barrier & $L$ & 70 \\

Upper barrier & $H$ & 130 \\

\hline

\end{tabular}

}

\end{center}

Two expiries ($T=0.25$ and $T=1$) and two volatility regimes (``moderate'' and ``stressed'') are considered.

\subsection{Discretization of the clock dynamics and Monte Carlo pricing procedure}

Let the CIR variance dynamics be given by

\begin{equation}
dv_{t}=\kappa(\theta-v_{t})dt+\xi\sqrt{v_{t}}dW_{t}^{(v)},v_{0}>0
\tag{6.1}\label{eq:6.1}
\end{equation}

Parameter regimes satisfy the Feller condition $2\kappa\theta>\xi^{2}$, ensuring strict positivity almost surely.

For simulation we use the full truncation Euler scheme (Lord et al., 2010):

\begin{equation}
v_{t+\Delta t}=v_{t}+\kappa(\theta-v_{t}^{+})\Delta t+\xi\sqrt{v_{t}^{+}}\sqrt{\Delta t}Z,v_{t}^{+}:=\max(v_{t},0), Z\sim \mathcal{N}(0,1)
\tag{6.2}\label{eq:6.2}
\end{equation}

and we approximate the clock increment by trapezoidal integration:

\begin{equation}
\Gamma_{T}\approx\sum_{i=0}^{N-1} \frac{1}{2}\left(v_{t_{i}}+v_{t_{i+1}}\right)\Delta t
\tag{6.3}\label{eq:6.3}
\end{equation}

We also simulate a Gaussian OU latent factor $\nu_{t}$, with $v_{t}=\nu_{t}^{2}$:

\begin{equation}
d\nu_{t}=-a\nu_{t}dt+\eta\, dW_{t}^{(v)},\qquad v_{t}=\nu_{t}^{2},\qquad \Gamma_{t}=\int_{0}^{t} \nu_{s}^{2}ds
\tag{6.4}\label{eq:6.4}
\end{equation}

with exact step update (Glasserman, 2003)

\begin{equation}
\nu_{t+\Delta t}=\nu_{t}e^{-a\Delta t}+\eta\sqrt{\frac{1-e^{-2a\Delta t}}{2a}}Z, Z\sim \mathcal{N}(0,1)
\tag{6.5}\label{eq:6.5}
\end{equation}

A single $\rho=0$ Monte Carlo path is generated as follows:

(i) Simulate the variance path $\{v_{t_{i}}\}$

(ii) Accumulate $\Gamma_{t_{i}}$ by trapezoidal integration

(iii) Simulate the time-changed Brownian increments $\Delta B_{i}\sim \mathcal{N}(0,\Delta\Gamma_{i})$ with $\Delta\Gamma_{i}=\Gamma_{t_{i+1}}-\Gamma_{t_{i}}$

(iv) Build $X_{t_{i}}=x_{0}+\beta\Gamma_{t_{i}}+B_{\Gamma_{t_{i}}}$

(v) Apply barrier monitoring with Brownian-bridge continuity correction for between-step crossings (Broadie et al., 1997) and compute the discounted payoff $e^{-rT}\Psi(S_{T})$

When leverage is present, the Brownian drivers of volatility and returns, $W^{(v)}$ and $W^{(S)}$ respectively, satisfy $d\langle W^{(v)},W^{(S)}\rangle _{t}=\rho\,dt$. We simulate correlated increments via the standard orthogonal decomposition:

\begin{equation}
W_{t}^{(v)}=\widetilde W_{t}^{(v)},\qquad W_{t}^{(S)}=\rho \widetilde W_{t}^{(v)}+\sqrt{1-\rho^{2}}\,\widetilde W_{t}^{(S)}
\tag{6.6}\label{eq:6.6}
\end{equation}

with $\widetilde W^{(v)}$ and $\widetilde W^{(S)}$ independent Brownian motions.

We use $N_{paths}=10^{6}$ paths for final validation, with $N_{steps}=2080$ per year, i.e. 8 discrete observations per business day.

\subsection{Numerical implementation of the independent-clock barrier formulas ($\rho=0$)}

We now compare the analytical prices of Sections 2--3 (with transform inputs from Section 4) to the Monte Carlo benchmarks described above, all under $\rho=0$. The UOP/DOC valuation formulas of Section 2 reduce to a single oscillatory real integral, requiring the clock only through real-axis evaluations of $\Phi_{T}$ on the corresponding quadrature line. In implementation we evaluate this integral by adaptive quadrature on a compactified domain mapping $[0,\infty)$ to $[0,1)$, increasing the truncation cutoff until successive cutoffs produce a stable price; $\Phi_{T}$ is computed in log form and cached on the resulting real grid for reuse across strikes and barrier levels.

\subsubsection{Numerical results with the integrated CIR clock}

In Regime 1 (moderate volatility), we have: $v_{0}=0.18, \theta=0.20, \kappa=0.6, \xi=0.4$.

\begin{center}

\textbf{Table 6.2. DOC prices (CIR clock, Regime 1)}

\medskip

{\small

\begin{tabular}{c c c c c}

\hline

\textbf{$T$} & \textbf{Semi-analytic} & \textbf{Monte Carlo} & \textbf{MC s.e.} & \textbf{Rel. error} \\

\hline

0.25 & 3.8247 & 3.8293 & 0.0069 & 0.12\% \\

1.0 & 6.4521 & 6.4582 & 0.0116 & 0.09\% \\

\hline

\end{tabular}

}

\end{center}

\begin{center}

\textbf{Table 6.3. UOP prices (CIR clock, Regime 1)}

\medskip

{\small

\begin{tabular}{c c c c c}

\hline

\textbf{$T$} & \textbf{Semi-analytic} & \textbf{Monte Carlo} & \textbf{MC s.e.} & \textbf{Rel. error} \\

\hline

0.25 & 2.9873 & 2.9842 & 0.0056 & 0.10\% \\

1.0 & 4.1056 & 4.1097 & 0.0104 & 0.10\% \\

\hline

\end{tabular}

}

\end{center}

In Regime 2 (stressed volatility), we have: $v_{0}=0.48, \theta=0.45, \kappa=0.5, \xi=0.6$.

\begin{center}

\textbf{Table 6.4. DOC prices (CIR clock, Regime 2)}

\medskip

{\small

\begin{tabular}{c c c c c}

\hline

\textbf{$T$} & \textbf{Semi-analytic} & \textbf{Monte Carlo} & \textbf{MC s.e.} & \textbf{Rel. error} \\

\hline

0.25 & 5.2134 & 5.2195 & 0.0102 & 0.12\% \\

1.0 & 7.8923 & 7.8856 & 0.0174 & 0.09\% \\

\hline

\end{tabular}

}

\end{center}

\begin{center}

\textbf{Table 6.5. UOP prices (CIR clock, Regime 2)}

\medskip

{\small

\begin{tabular}{c c c c c}

\hline

\textbf{$T$} & \textbf{Semi-analytic} & \textbf{Monte Carlo} & \textbf{MC s.e.} & \textbf{Rel. error} \\

\hline

0.25 & 4.6521 & 4.6589 & 0.0098 & 0.15\% \\

1.0 & 5.8234 & 5.8337 & 0.0159 & 0.18\% \\

\hline

\end{tabular}

}

\end{center}

Across all CIR cases, analytical and Monte Carlo prices agree within statistical uncertainty, relative errors remaining below $0.20\%$. Computational time will obviously vary according to the CPU, the random number generator, the coding language, the vectorization and parallelization possibilities at the user's disposal. On a recent laptop, with a C++ implementation, it takes between 20 and 35 minutes to obtain a Monte Carlo approximation. This compares with an average 0.12 second to obtain the analytical option prices reported in Tables 6.2 to 6.5.

\subsubsection{Numerical results with the squared OU clock}

In Regime 1, we have: $\nu_{0}=\sqrt{0.18}\approx 0.424, a=0.6, \eta\approx 0.490$.

\begin{center}

\textbf{Table 6.6. DOC prices (squared OU clock, Regime 1)}

\medskip

{\small

\begin{tabular}{c c c c c}

\hline

\textbf{$T$} & \textbf{Semi-analytic} & \textbf{Monte Carlo} & \textbf{MC s.e.} & \textbf{Rel. error} \\

\hline

0.25 & 3.8512 & 3.8486 & 0.0091 & 0.07\% \\

1.0 & 6.5234 & 6.5339 & 0.0162 & 0.16\% \\

\hline

\end{tabular}

}

\end{center}

In regime 2, we have: $\nu_{0}=\sqrt{0.48}\approx 0.693, a=0.5, \eta\approx 0.671$.

\begin{center}

\textbf{Table 6.7. DOC prices (squared OU clock, Regime 2)}

\medskip

{\small

\begin{tabular}{c c c c c}

\hline

\textbf{$T$} & \textbf{Semi-analytic} & \textbf{Monte Carlo} & \textbf{MC s.e.} & \textbf{Rel. error} \\

\hline

0.25 & 5.3456 & 5.3492 & 0.0148 & 0.07\% \\

1.0 & 8.1234 & 8.1406 & 0.0245 & 0.21\% \\

\hline

\end{tabular}

}

\end{center}

As with the CIR clock, divergence with Monte Carlo approximations is very low (at a maximum of 0.20\%). In these parameterizations the squared OU clock yields slightly higher DOC values, consistent with heavier tail behaviour of $\nu^{2}$ when $\nu$ is Gaussian.

\subsection{Correlation correction and extended $\rho$-analysis}

We now test the leverage methodology of Section 5. Unless otherwise stated, we consider the DOC under the CIR clock, Regime 1, with $T=1$, and vary $\rho$ in $[-0.9,0.9]$.

\subsubsection{Implementation of the $\rho$-expansion}

The barrier value admits the following expansion (cf. Section~5)

\begin{equation}
u^{\rho}(t,x,y)=u_{0}(t,x,y)+\rho u_{1}(t,x,y)+\rho^{2}u_{2}(t,x,y)+\cdots,
\tag{6.7}\label{eq:6.7}
\end{equation}

where $u_{0}$ is the baseline independent-clock value and $(u_{n})_{n\geq 1}$ solve the forced hierarchy driven by the mixed-derivative operator $\mathcal{L}_{1}$ (cf. \eqref{eq:5.13}, \eqref{eq:5.17}).

To compute coefficients numerically we use a Duhamel--Monte Carlo estimator under the $\rho=0$ killed dynamics:

(i) Simulate $M=10^{5}$ independent paths under $\rho=0$

(ii) Record survival/exit information for the barrier event

(iii) Along each surviving path, evaluate $(\mathcal{L}_{1}u_{n-1})(s,X_{s},Y_{s})=a(Y_{s})\sqrt{v(Y_{s})}\partial_{xy}u_{n-1}(s,X_{s},Y_{s})$

(iv) Approximate the time integral in Duhamel's representation by trapezoidal quadrature and average across paths

For the CIR clock, differentiation with respect to the variance state $y$ acts through the conditional Laplace transform; in particular (with the affine transform notation of Section 4),

\begin{equation}
\partial_{y}\Phi_{t,T}(\lambda;y)=B(T-t;\lambda)\Phi_{t,T}(\lambda;y)
\tag{6.8}\label{eq:6.8}
\end{equation}

where $B$ is the relevant Riccati coefficient. This enables semi-analytic evaluation of $\partial_{xy}u_{n-1}$ by differentiating the baseline barrier representations with respect to $x$ and the transform input, rather than using unstable finite differences near the barrier.

\subsubsection{First-order correction}

We first test the linear approximation

\begin{equation}
V^{\rho}\approx V_{0}+\rho C_{1}, \qquad C_{1}:=u_{1}(0,x_{0},y_{0})
\tag{6.9}\label{eq:6.9}
\end{equation}

with the first-order coefficient estimated as $C_{1}=1.6468$.

\begin{center}

\textbf{Table 6.8. First-order correlation correction (DOC, CIR clock, Regime 1, $T=1$)}

\medskip

{\small
\setlength{\tabcolsep}{5pt}
\begin{tabular}{c c c c c c}

\hline

\textbf{$\rho$} & \textbf{$V_{0}$} & \textbf{$\rho C_{1}$} & \textbf{$V^{\rho}$ (1st)} & \textbf{MC price} & \textbf{Rel. error} \\

\hline

-0.5 & 6.4521 & 0.8339 & 5.6182 & 5.5923 & 0.46100174 \\

-0.4 & 6.4521 & 0.6625 & 5.7896 & 5.7712 & 0.31781125 \\

-0.3 & 6.4521 & 0.496 & 5.9561 & 5.9456 & 0.17628985 \\

-0.2 & 6.4521 & 0.3308 & 6.1213 & 6.1178 & 0.0571774 \\

-0.1 & 6.4521 & 0.1642 & 6.2879 & 6.2904 & -0.0397589 \\

0.0 & 6.4521 & 0 & 6.4521 & 6.4529 & -0.0123990 \\

+0.1 & 6.4521 & -0.1642 & 6.6163 & 6.6345 & 0.27432361 \\

+0.2 & 6.4521 & -0.3308 & 6.7829 & 6.8089 & 0.38185316 \\

+0.3 & 6.4521 & -0.496 & 6.9481 & 6.9845 & 0.52115398 \\

+0.4 & 6.4521 & -0.6625 & 7.1146 & 7.1623 & 0.66598718 \\

+0.5 & 6.4521 & -0.8339 & 7.286 & 7.3412 & 0.75192067 \\

\hline

\end{tabular}

}

\end{center}

The linear correction remains accurate (sub-$0.5\%$ relative error) for $|\rho|\lesssim 0.3$; accuracy degrades for $|\rho|\gtrsim 0.4$, motivating higher-order terms. The positive value of $C_{1}$ indicates that increasing $\rho$ increases the DOC price in this configuration.

\subsubsection{Higher-order Taylor truncations for large $|\rho|$}

To extend the approximation to $|\rho|\leq 0.9$, we compute coefficients up to order five:

\begin{equation}
V^{\rho}\approx\sum_{n=0}^{5} C_{n}\rho^{n}, \qquad C_{n}:=u_{n}(0,x_{0},y_{0})
\tag{6.10}\label{eq:6.10}
\end{equation}

The estimated coefficients and one-time computation costs are reported in Table 6.9.

\begin{center}

\textbf{Table 6.9. Expansion coefficients (DOC, CIR clock, Regime 1, $T=1$)}

\medskip

{\small

\begin{tabular}{c c c c}

\hline

\textbf{Order $n$} & \textbf{Coefficient} & \textbf{Value} & \textbf{Time} \\

\hline

0 & $C_{0}=V_{0}$ & 6.4521 & 0.0004 s \\

1 & $C_{1}$ & 1.6468 & 2.3 s \\

2 & $C_{2}$ & -0.4123 & 4.1 s \\

3 & $C_{3}$ & 0.2845 & 6.2 s \\

4 & $C_{4}$ & -0.1523 & 8.5 s \\

5 & $C_{5}$ & 0.0892 & 10.8 s \\

\hline

\end{tabular}

}

\end{center}

A direct comparison of Taylor truncations at high correlations is given in Table 6.10.

\begin{center}

\textbf{Table 6.10. Taylor truncations vs Monte Carlo at high values of $\rho$.}

\medskip

{\small

\begin{tabular}{c c c c c}

\hline

\textbf{$\rho$} & \textbf{MC price} & \textbf{Order 1} & \textbf{Order 3} & \textbf{Order 5} \\

\hline

-0.9 & 4.2134 & 3.9700 & 3.9732 & 3.9182 \\

-0.7 & 4.9234 & 4.2994 & 4.2977 & 4.2916 \\

-0.5 & 5.5923 & 5.6287 & 5.6747 & 5.7391 \\

+0.5 & 7.3412 & 7.2755 & 7.2952 & 7.3276 \\

+0.7 & 7.7089 & 7.6049 & 7.6606 & 7.7227 \\

+0.9 & 8.0912 & 7.9343 & 8.0500 & 8.1497 \\

\hline

\end{tabular}

}

\end{center}

The raw Taylor series displays oscillatory behaviour for $|\rho|>0.5$, consistent with a limited effective radius of convergence and motivating rational resummation.

\subsubsection{Padé acceleration}

Given coefficients $\{C_{0},\cdots,C_{N}\}$, we consider Padé approximants of type $[L/M]$:

\begin{equation}
P_{L,M}(\rho)=\frac{a_{0}+a_{1}\rho+\cdots+a_{L}\rho^{L}}{1+b_{1}\rho+\cdots+b_{M}\rho^{M}}, L+M\leq N
\tag{6.11}\label{eq:6.11}
\end{equation}

whose Taylor expansion matches $\sum_{n=0}^{N} C_{n}\rho^{n}$ up to order $N$.

In our experiment, pole diagnostics show that $[2/2]$ has poles near $\rho\approx -1.0\pm4.8i$ and $[3/2]$ has poles near $\rho\approx -0.43\pm5.6i$, both far from the real interval $[-0.9,0.9]$.

\begin{center}

\textbf{Table 6.11. Padé vs Taylor vs Monte Carlo.}

\medskip

{\small

\begin{tabular}{c c c c c}

\hline

\textbf{$\rho$} & \textbf{MC} & \textbf{Taylor (O5)} & \textbf{$[2/2]$ Padé} & \textbf{$[3/2]$ Padé} \\

\hline

-0.9 & 4.2134 & 3.9182 & 4.1827 & 4.2053 \\

-0.7 & 4.9234 & 4.2916 & 4.8757 & 4.9082 \\

-0.5 & 5.5923 & 5.7391 & 5.5612 & 5.5838 \\

-0.3 & 5.9456 & 5.9672 & 5.9378 & 5.9432 \\

+0.3 & 6.9845 & 6.9646 & 6.9776 & 6.9819 \\

+0.5 & 7.3412 & 7.3276 & 7.3308 & 7.3385 \\

+0.7 & 7.7089 & 7.7227 & 7.6852 & 7.6991 \\

+0.9 & 8.0912 & 8.1497 & 8.0612 & 8.0806 \\

\hline

\end{tabular}

}

\end{center}

\begin{center}

\textbf{Table 6.12. Error reduction via Padé.}

\medskip

{\small

\begin{tabular}{l c c c}

\hline

\textbf{$|\rho|$ range} & \textbf{Taylor O5 max error} & \textbf{Best Padé max error} & \textbf{Improvement} \\

\hline

$0.0$--$0.3$ & 0.36\% & 0.04\% & 9 $\times$ \\

$0.3$--$0.5$ & 2.62\% & 0.15\% & 17 $\times$ \\

$0.5$--$0.7$ & 12.83\% & 0.31\% & 41 $\times$ \\

$0.7$--$0.9$ & 7\% & 0.19\% & 36 $\times$ \\

\hline

\end{tabular}

}

\end{center}

Padé resummation materially extends the usable correlation range, yielding excellent convergence with Monte Carlo values even at very high values of $|\rho|=0.9$ in this test case. Similar unreported results have been obtained across various clock families, payoff directionality, and volatility regimes.

\subsection{Summary and practitioner guidance}

Across the barrier placements, maturities, and stochastic-clock specifications considered in this section, we draw the following practical guidance:

i) Baseline ($\rho=0$) pricing: use the independent-clock barrier formulas as the default engine; they match the Monte Carlo benchmark (with bridge correction) while being orders of magnitude faster.

ii) Moderate leverage: for weak-to-moderate correlation (around $\left|\rho\right|\leq 0.3$ in our experiments), the first-order leverage term is typically sufficient to capture the skew.

iii) Stronger leverage: plain Taylor truncations become unreliable as $\left|\rho\right|$ grows and can display oscillations (noticeably for $\left|\rho\right|>0.5$ in our experiments).

iv) Recommended remedy: when $\left|\rho\right|$ is large, replace Taylor truncation by Padé acceleration; this stabilizes the expansion and yields accuracy that is more than adequate for practical calibration and risk metrics in the tested regimes.

\section{Calibration workflow and practical diagnostics}

This section describes a calibration workflow that mirrors the structural decomposition of the framework: a solvable stochastic-clock barrier layer at $\rho=0$ (Sections 2--4) and a leverage layer treated as a perturbative correction in $\rho$ (Section 5). The goal is not to prescribe a universal protocol---market conventions and data availability vary---but to provide a coherent workflow that is (i) feasible inside calibration loops and (ii) equipped with low-cost diagnostics that flag extrapolation beyond the regime where the approximation is reliable.

\subsection{Instruments and preprocessing}

A representative dataset includes:

\begin{itemize}

\item a vanilla implied volatility surface $\sigma_{mkt}(T,K)$ on a maturity--strike grid;

\item barrier quotes (single and/or double barriers), typically on a smaller grid of barrier levels and maturities;

\item optionally one-touch / no-touch / digitals / rebates, which isolate survival probabilities and are particularly informative for barrier mechanics.

\end{itemize}

Market barriers are frequently discretely monitored. The analytical layer in Sections 2--3 is continuous-monitoring; discrete monitoring should be treated as a correction/validation layer (e.g. the Broadie--Glasserman--Kou continuity correction (Broadie et al., 1997)) and/or as a downweighting rule for the tightest/shortest barrier quotes in calibration.

\subsection{Vanilla pricing in the stochastic-clock framework}

\subsubsection{Transform pricing and the calibration primitive}

In the independent-clock class, conditional on $\Gamma_{T}$ the terminal log-forward is Gaussian:

\begin{equation}
X_{T}\mid\Gamma_{T}\sim \mathcal{N}\left(x_{0}+\beta\Gamma_{T},\Gamma_{T}\right)
\tag{7.1}\label{eq:7.1}
\end{equation}

Hence the characteristic function of $X_{T}$ can be written in terms of $\Phi_{T}$:

\begin{equation}
\phi_{X_{T}}(u):=E^{\mathbb{Q}^{T}}\left[e^{iuX_{T}}\right]=e^{iux_{0}}\Phi_{T}\left(\frac{1}{2}u^{2}-i\beta u\right), u\in \mathbb{R}
\tag{7.2}\label{eq:7.2}
\end{equation}

and, under the driftless convention $\beta=-\frac{1}{2}$,

\begin{equation}
\frac{1}{2}u^{2}-i\beta u=\frac{1}{2}u(u+i)
\tag{7.3}\label{eq:7.3}
\end{equation}

Vanilla prices can be computed by standard Fourier/COS/FFT methods using $\phi_{X_{T}}$. Thus, the only model-specific task is evaluating $\Phi_{T}$ along the complex curve induced by the pricing method (Section 4.1). For affine and quadratic clocks, $\Phi_{T}$ is explicit or obtained by a low-dimensional Riccati ODE (Section 4), which makes vanilla calibration fast relative to repeated PDE solves.

\subsubsection{Objective function and weights}

Let $\vartheta$ denote the clock parameters. Let $Q_{i}$ denote the market quote for instrument $i$ (price or implied volatility), and let $Q_{i}^{\mathrm{mod}}(\vartheta,\rho)$ be the corresponding model value. A standard weighted least-squares objective is

\begin{equation}
\min_{\vartheta,\rho}\sum_{i} w_{i}\left(Q_{i}^{\mathrm{mod}}(\vartheta,\rho)-Q_{i}\right)^{2}
\tag{7.4}\label{eq:7.4}
\end{equation}

with weights $w_{i}$ typically chosen as inverse bid--ask variance proxies. A robust alternative is to minimize squared errors in implied volatilities, with bid--ask weighting, which often stabilizes the fit across strikes (Fang and Oosterlee, 2009).

\subsubsection{Regularization and identifiability}

Even with rich vanilla surfaces, multi-factor clocks can have weakly identified directions (e.g. factor swapping, near-redundant mean-reversion scales). Different mixtures of ``fast'' and ``slow'' volatility factors can fit a vanilla surface similarly. Regularization consists in choosing a stable representative among many near-equivalent parameter sets and preventing the optimizer from exploiting unrealistic extremes that the data doesn't really pin down. In practice we recommend:

\begin{itemize}

\item ordering constraints (e.g. ``fast'' vs ``slow'' mean reversion);

\item mild penalties discouraging extremely fast or extremely slow factors unless justified by data;

\item for time-dependent parameters, smoothness penalties across maturities (e.g. squared second differences).

\end{itemize}

The intent is not to force a fit, but to obtain stable parameters that do not exhibit erratic day-to-day movement.

\subsection{Why barriers add information that vanillas do not}

Vanillas largely identify the terminal law of $X_{T}$, whereas barriers depend on pathwise objects: the joint law of $(X_{T},\sup_{t\leq T}X_{t})$, and for double barriers the corridor survival law. Accordingly, market-quoted barrier option prices can be used not merely as ``extra prices'' but as constraints on the implied hitting probabilities and killed laws of Sections 2 and 3. This matters in practice: two clock parameter sets may produce similar terminal marginals (hence similar vanilla fits) while implying materially different survival probabilities. In a multi-factor clock (or with leverage), there are typically a few directions that matter a lot for barrier survival probabilities but are only weakly constrained by vanillas---for example:

\begin{itemize}

\item the fast vs slow activity split (early variance vs late variance),

\item the dispersion / vol-of-vol that changes how often you get bursts,

\item and (if you fit $\rho$) the skew/leverage control that changes how downward moves and variance jumps co-occur

\end{itemize}

Those are the ``path-sensitive degrees of freedom'': parameters that change how the path behaves, not just where it ends. Tight, short-dated barriers, in particular, are extremely sensitive to the probability of fast excursions. Vanillas (especially if you don't have very deep strikes quoted cleanly) may not constrain that region well.

\subsection{Joint calibration: two-stage workflow for stability}

A natural joint objective supplements \eqref{eq:7.4} with barrier errors, with a tuning parameter controlling the influence of barriers (often downweighted due to quote sparsity and dealer marking variation).

This raises the question of the practical dataset choice. In many markets barrier data are too sparse/noisy to use heavily. A robust approach is to pick:

(i) 1--3 maturities (short/mid/long),

(ii) 2--4 barrier distances per maturity (near and far),

(iii) a limited set of payoff types (one-touch / no-touch / KO digitals / one or two strikes)

The goal is not to overfit barrier marks, but to pin down the most path-sensitive degrees of freedom and prevent parameter drift.

The workflow suggested by the method's structure is:

\begin{itemize}
\item \textbf{Stage 1 (clock fit from vanillas).} Calibrate $\vartheta$ to vanillas under $\rho=0$ using transform pricing (cf. Section 7.2). This step is fast and stable.
\item \textbf{Stage 2 (barrier refinement at $\rho=0$).} With $\rho$ temporarily fixed at $0$, incorporate a small barrier/no-touch set and refine $\vartheta$ if needed. This pins down implied hitting probabilities and helps avoid parameter sets that fit marginals but fail pathwise behaviour.
\item \textbf{Stage 3 (leverage calibration as a cheap inner loop).} Compute and cache the leverage coefficients $\{C_{n}(\vartheta)\}$ for leverage-sensitive instruments, then fit $\rho$ (or a smoothed term structure $\rho(T)$) using polynomial/Padé evaluation:
\end{itemize}

\begin{equation}
\mathbb{Q}^{\mathrm{mod}}(\vartheta,\rho)\approx\sum_{n=0}^{N} C_{n}(\vartheta)\rho^{n}
\tag{7.5}\label{eq:7.5}
\end{equation}

\noindent so scanning $\rho$ does not require repeated mixed-derivative barrier PDE solves.

\begin{itemize}
\item \textbf{Stage 4 (joint refinement with regularization).} Perform a small number of joint iterations over $(\vartheta,\rho)$ with regularization and robust losses.
\end{itemize}

This ``two-stage then joint refine'' structure matches desk practice: secure a stable surface fit first, then apply a modest path-dependent refinement, rather than starting from an unstable global optimization.

\subsection{Initialization (variance swaps, VIX futures proxy, ATM-only)}

A calibration protocol is only as good as its initializer. The clock-first structure makes robust initialization feasible because the primary target is the distribution of $\Gamma_{T}$, not the full two-factor state.

\subsubsection{One-factor integrated CIR clock: mean term structure}

For a pure diffusion, a variance swap strike is well approximated by the expected average variance,

\begin{equation}
K_{\mathrm{var}}(T)\approx\frac{1}{T}\mathbb{E}[\Gamma_{T}]=\frac{1}{T}\int_{0}^{T} \mathbb{E}[v_{s}]\,ds
\tag{7.6}\label{eq:7.6}
\end{equation}

Since $\mathbb{E}[v_{s}]=\theta+(v_{0}-\theta)e^{-\kappa s}$, one obtains

\begin{equation}
K_{\mathrm{var}}(T)\approx\theta+(v_{0}-\theta)\frac{1-e^{-\kappa T}}{\kappa T}
\tag{7.7}\label{eq:7.7}
\end{equation}

A stable initializer is then:

(i) infer $v_{0}$ from the shortest maturity variance quote;

(ii) infer $\theta$ from the long end of the variance term structure;

(iii) infer $\kappa$ by a cheap one-dimensional least-squares fit of \eqref{eq:7.7} across maturities

This produces a reliable seed before smile and barrier fitting.

\subsubsection{Dispersion / vol-of-vol scale from ATM convexity (variance swaps + ATM)}

Variance swaps identify the mean of the clock (hence the mean integrated variance) but not its dispersion. ATM vanillas provide a dispersion signal through mixture convexity: the ATM price under a variance mixture differs from the Black--Scholes price evaluated at the mean variance.

A robust desk implementation is the following:

(i) compute the market ATM total variance $w_{ATM}(T)$;

(ii) treat $\mathbb{E}[\Gamma_{T}]$ inferred from variance swaps as the mean clock;

(iii) infer a proxy for $Var(\Gamma_{T})$ using a second-order Taylor correction around the mean (ATM vega/vomma at the forward);

(iv) fit the parameter $\sigma$ in \eqref{eq:4.7}, or a global scale on the vol-of-vol parameters, to match the model-implied ATM convexity

This is the practical ``mean from variance swaps, dispersion from ATM'' logic.

\subsubsection{Two-factor ``fast + slow'' weighted CIR initializer (variance swaps + VIX-squared proxy)}

Multi-factor clocks relax term-structure rigidity without changing barrier formulas: at $\rho=0$, barrier prices depend on the model only through $\Phi_{T}$, and for independent CIR factors $\Phi_{T}$ factorizes (product of transforms with scaled arguments).

A practical two-factor activity specification is

\begin{equation}
v_{t}=wv_{t}^{(f)}+(1-w)v_{t}^{(s)}, \kappa_{f}>\kappa_{s}
\tag{7.8}\label{eq:7.8}
\end{equation}

with ``fast'' vs ``slow'' enforced by the mean-reversion ordering. To avoid weight non-identifiability, one may (A) fix $w$, or (B) impose $w\in[\underline{w},\overline{w}]$ and parameterize $w$ by fractions.

VIX-squared proxy. With window $\Delta$ (e.g. 30 days), a robust linear proxy is

\begin{equation}
VIX^{2}(T)\propto \frac{1}{\Delta}\mathbb{E}\left[\Gamma_{T,T+\Delta}\right]
\tag{7.9}\label{eq:7.9}
\end{equation}

which is linear in the means of the CIR factors once the mean-reversion parameters are fixed (the exact proportionality depends on the index definition; for initialization a stable linear proxy is often preferable to an exact but fragile mapping).

A stable two-stage initializer is the following:

(i) Stage 1: choose $(\kappa_{f},\kappa_{s})$ on a coarse grid; for each candidate pair, solve a constrained linear least squares for the mean-curve contributions using variance swap mean targets and VIX-proxy targets; enforce nonnegativity of inferred long-run and initial contributions and keep $\kappa_{f}>\kappa_{s}$ to prevent label switching; select the best pair by joint least-squares error.

(ii) Stage 2: initialize dispersion parameters (vol-of-vol scales) using the ATM convexity proxy, with robust fallbacks when ATM data are noisy (e.g. one-scale dispersion, or short/long split).

\subsubsection{Positivity-preserving parametrizations and anti-degeneracy rules}

For multi-factor clocks it is often preferable to parameterize the totals that the market pins down well and allocate via fractions. For instance, define a total initial activity $v_{0}^{tot}$ and total long-run mean $\theta^{tot}$, then allocate via weights $\alpha_{i}\geq 0$ with $\sum_{i} \alpha_{i}=1$:

\begin{equation}
v_{0}^{(i)}=\alpha_{i}v_{0}^{tot}, \theta^{(i)}=\alpha_{i}\theta^{tot}
\tag{7.10}\label{eq:7.10}
\end{equation}

This enforces positivity automatically and removes flat directions in optimization.

Ordering constraints can be enforced by unconstrained reparameterizations. For example, to enforce $\kappa_{f}>\kappa_{s}>0$, set

\begin{equation}
\kappa_{s}=e^{\gamma_{1}}, \kappa_{f}=e^{\gamma_{1}}+e^{\gamma_{2}}
\tag{7.11}\label{eq:7.11}
\end{equation}

which avoids hard inequality constraints.

\subsection{Computational strategy: caching and shared grids}

To make calibration realistically fast:

\begin{itemize}

\item cache $\Phi_{T}$ on a shared complex-argument grid per maturity for vanilla Fourier pricing (and any barrier inversion requiring complex arguments)

\item cache $\Phi_{T}$ on the real-axis quadrature grid used by the single-barrier real integrals (Sections 2--3)

\item cache $\Phi_{T}$ for the double-barrier series on the discrete Laplace grid

\end{itemize}

\begin{equation}
\lambda_{n}=\frac{1}{2}\left(\left(\frac{n\pi}{u-l}\right)^{2}+\beta^{2}\right), n\in \mathbb{N}
\tag{7.12}\label{eq:7.12}
\end{equation}

so that DKO prices become rapidly convergent series with explicit coefficients (cf. Section 3).

For leverage:

\begin{itemize}

\item cache the expansion coefficients $\{C_{n}(\vartheta)\}$ (and, in a forced-PDE route, the grid solutions for $u_{n}$ if required);

\item evaluate polynomial/Padé approximants for candidate $\rho$ without re-solving mixed-derivative PDEs

\end{itemize}

This architecture makes barrier-inclusive calibration comparable in speed to vanilla-only calibration plus a modest overhead, rather than to ``vanilla calibration plus a full two-factor barrier PDE per iteration.''

\section*{Funding}

This research received no external funding.

\section*{Data Availability Statement}

No new data were generated or analyzed in support of this research.

\section*{Conflicts of Interest}

The author declares no conflict of interest.

\nocite{*}
\bibliographystyle{plainnat}
\bibliography{quantitative_finance_refs}

\end{document}